\documentclass[aoas]{imsart}

\RequirePackage{amsthm,amsmath,amsfonts,amssymb}
\RequirePackage[authoryear]{natbib}
\RequirePackage[colorlinks,citecolor=blue,urlcolor=blue]{hyperref}
\RequirePackage{graphicx}

\usepackage{latexsym,epsfig,epstopdf,amscd,multirow,psfrag,paralist,dsfont,url,color,caption,subcaption}
\usepackage[titletoc]{appendix}
\usepackage[american]{babel}

\usepackage{chngcntr}
\usepackage{float}

\startlocaldefs

\usepackage[ruled,vlined]{algorithm2e}
\SetKwInput{KwInput}{Input}                
\SetKwInput{KwOutput}{Output}              
\SetKwInput{KwData}{Data}              

\newcommand{\thetavec}{{\boldsymbol{\theta}}}
\newcommand{\varthetavec}{{\boldsymbol{\vartheta}}}

\newcommand{\uvec}{{\boldsymbol{u}}}
\newcommand{\vvec}{{\boldsymbol{v}}}
\newcommand{\wvec}{{\boldsymbol{w}}}

\newcommand{\zvec}{{\boldsymbol{z}}}

\newcommand{\alphavec}{\boldsymbol{\alpha}}
\newcommand{\betavec}{{\boldsymbol{\beta}}}
\newcommand{\phivec}{{\boldsymbol{\phi}}}
\newcommand{\psivec}{{\boldsymbol{\psi}}}
\newcommand{\gammavec}{{\boldsymbol{\gamma}}}
\newcommand{\Psivec}{{\boldsymbol{\Psi}}}

\newcommand{\degreeC}{\ensuremath{^\circ}\text{C}}

\newcommand{\N}{\mathcal{N}}

\theoremstyle{plain}

\theoremstyle{remark}

\endlocaldefs

\begin{document}

\begin{frontmatter}
\title{Reliability Study of Battery Lives: A Functional Degradation Analysis Approach}
\runtitle{Functional Degradation Analysis of Battery Lives}

\begin{aug}
\author[A]{\fnms{Youngjin}~\snm{Cho}\ead[label=e1]{youngjin@vt.edu}},
\author[B]{\fnms{Quyen}~\snm{Do}\ead[label=e2]{DoQN@corning.com}},
\author[A]{\fnms{Pang}~\snm{Du}\ead[label=e3]{pangdu@vt.edu}\orcid{0000-0003-1365-4831}}
\and
\author[A]{\fnms{Yili}~\snm{Hong}\ead[label=e4]{yilihong@vt.edu}}
\address[A]{Department of Statistics,
Virginia Tech\printead[presep={,\ }]{e1,e3,e4}}
\address[B]{Corning Inc.\printead[presep={,\ }]{e2}}

\end{aug}

\begin{abstract}
Renewable energy is critical for combating climate change, whose first step is the storage of electricity generated from renewable energy sources. Li-ion batteries are a popular kind of storage units. Their continuous usage through charge-discharge cycles eventually leads to degradation. This can be visualized by plotting voltage discharge curves (VDCs) over discharge cycles. Studies of battery degradation have mostly concentrated on modeling degradation through one scalar measurement summarizing each VDC. Such simplification of curves can lead to inaccurate predictive models. Here we analyze the degradation of rechargeable Li-ion batteries from a NASA data set through modeling and predicting their full VDCs. With techniques from longitudinal and functional data analysis, we propose a new two-step predictive modeling procedure for functional responses residing on heterogeneous domains. We first predict the shapes and domain end points of VDCs using functional regression models. Then we integrate these predictions to perform a degradation analysis. Our functional approach allows the incorporation of usage information, produces predictions in a curve form, and thus provides flexibility in the assessment of battery degradation. Through extensive simulation studies and cross-validated data analysis, our approach demonstrates better prediction than the existing approach of modeling degradation directly with aggregated data.
\end{abstract}


\end{frontmatter}

	\section{Introduction}\label{se:intro}
	Renewable energy plays an important role in global energy security and global warming control \citep{olabi2022renewable}. Commonly used renewable energy technologies include solar power technology, wind turbines, biomass power generation technology and tidal energy technology \citep{bull2001renewable,panwar2011role}. Once electricity is generated from those renewable energy technologies, energy storage batteries are needed for energy to be stored when available and released when needed. Li-ion batteries are widely used for energy storage and one significant application is on electric vehicles \citep{diouf2015potential,castelvecchi2021electric}.
	
	An important issue in these applications is the assessment of battery life under a certain use condition. Accurate assessment of battery lives is critical for large-scale deployments of renewable technology. This motivates battery life studies like the one reported in \citet{saha2007battery}, where the question of interest is how many charge-discharge cycles a battery can sustain before it reaches a failure threshold in terms of capacity loss.
	Later, \citet{saha2009pf} proposed to predict battery capacity under the particle filtering (PF) framework. They also developed a Rao-Blackwellized PF approach in \citet{saha2008rbpf} to reduce uncertainty on capacity prediction. A comparison of their methods is available in \citet{saha2009comparison}. Since then various statistical and machine learning methods have been proposed for the battery capacity prediction problem. Examples of statistical models include stochastic models using autoregressive processes \citep{liu2012nlar} or Wiener processes \citep{tang2014weiner,shen2018wiener}, and Gaussian process regression \citep{liu2012gpr,liu2013gpfr,he2015gpr,richardson2017gpr,tagade2020deep}. Examples of machine learning techniques include na\"{i}ve Bayes \citep{ng2014naive}, support vector machine \citep{patil2015svm}, relevance vector machine \citep{liu2015rvm}, genetic algorithms \citep{yu2022ega}, and neural networks \citep{sbarufatti2017rbnn,nascimento2021hybrid}.  \citet{liu2014fusion} provided a fusion method integrating regularized particle filtering and autoregressive models. A comprehensive review of these methods can be found in \citet{martin2022comparison}.
 Most of the aforementioned work focused only on a small number of batteries ($<5$) under the same experimental conditions. Other studies, such as \citet{tagade2020deep}, \citet{ng2014naive}, and \citet{patil2015svm}, built up predictive models for more batteries, but focused only on the prediction of battery capacity.  \citet{sbarufatti2017rbnn} considered the prediction within a single cycle. \citet{nascimento2021hybrid} studied 12 batteries combining recurrent neural networks with physics-based knowledge on battery aging to predict discharge curves for a battery. However, even these more recent studies did not take into account different experimental conditions that could affect the aging behaviors of batteries.
	
	Even though there is vast literature on the modeling and prediction of Li-ion battery lives, most existing analyses use a variety of techniques to reduce the discharge curves to simple numerical summaries and then apply traditional statistical methods to these summaries. They do not take full advantage of the whole discharge curves, each of which gives much more details than a simple summary number, such as the area under the curve, on how the battery performed in a cycle. Figure \ref{fig:discharge_curves} illustrates the voltage discharge curves (VDCs) of two batteries in the data set of interest. An aggregation summary such as area under the curve does represent a type of degradation measure, but there is clearly the risk of information loss since the whole curve is not taken into account. This motivates us to adopt the functional data analysis (FDA) approach to model and predict battery lives. Ever since its introduction by \citet{fda1997}, FDA has found many extensions to various areas in statistics; see, e.g., the reviews by \citet{Wang2016}, \citet{recentdev2019}, and \citet{fda_jmva}. Under the FDA framework, each curve is viewed as a functional observation.  Since the VDCs for the same battery are naturally correlated, random effects are required in the model to incorporate such within-battery correlation. Hence, we model the VDCs through a functional linear mixed-effects model proposed in \citet{liu2017estimating}. Some other options include \citet{guo02} and \citet{zhu19}. As a modern technique, the FDA approach has been rarely used 
	in the area of degradation modeling. One exception is \citet{zhou:11}, which focused on modeling possibly incomplete degradation profiles due to sparse or fragmented signals. Their method is a combination of local polynomial smoothing,  functional principal component analysis (FPCA), and empirical Bayes. They assume that, incomplete or not, all the degradation profiles live on the same time domain. This makes it difficult to easily extend to study the VDCs here, which live on heterogeneous domains with the domain end points carrying useful information as well.

 A popular approach to degradation analysis is the general path model (GPM), for which a good reference is \citet{meeker2021}, covering the basics of GPM modeling. \citet{xie2015} provide a comprehensive review on degradation data models. Recent development in degradation models focuses more on incorporating time varying covariates \citep{hong2015} and on describing dependence structure of multiple degradation characteristics \citep{fang2023}. However, there is no existing work that directly models functional degradation data as in this paper.
	
	Modeling functional curves with covariate information falls under the umbrella of functional response models. Existing functional response models require a common domain for the functional responses. However, in our battery data, each VDC has its distinctive domain, ranging from the start of the discharge cycle to the so-called {\it end of discharge} (EOD) time \citep{saha2009pf}. Therefore, we first introduce a scaled version of VDCs where each VDC is scaled by its EOD to lie on the common domain of $[0,1]$. Then we consider a model consisting of two parts. One part models the EODs and the other models the scaled VDCs. For the predictors in the models, experimental conditions are the natural covariates to include. The cycle number is also necessary for the modeling of the degradation patterns. Some other covariates for consideration include the resting time between cycles, the EOD of the previous cycle, and the scaled VDC itself. To model the scaled VDCs, we first conduct an FPCA to represent each curve by its projected scores onto the leading functional principal components (FPCs). Then a multivariate mixed-effects model is fitted with the random effects introduced to incorporate the within-battery correlations between VDCs. For the EODs, we consider two versions of mixed-effects models, with or without the inclusion of the functional covariate of scaled VDCs. Our degradation analyses, with three versions, are all based on these functional models of the scaled curves and EODs. The first version is the standard general path model where degradation amounts computed from fitted VDCs at the observed cycles (or training cycles) are the only inputs to fit the prediction model for degradation amounts. The other two versions, corresponding to the two versions of EOD models, would use the functional models to predict scaled VDCs and EODs first and then use them to calculate the predicted degradation amounts.
	
	Through extensive simulations and comprehensive analysis of the battery data, we show the great potential of an FDA approach to degradation analysis of batteries. In particular, it has the following  distinct contributions to the degradation analysis field: (i) it makes full use of the functional forms of the original data to ensure a minimal information loss; (ii) the FDA framework also allows an easy incorporation of covariates such as experimental conditions, which are critical factors for degradation but have been largely ignored by the existing battery degradation studies; (iii) it is demonstrated in our numerical studies to have much better prediction performance than the traditional aggregation methods, which is essential in degradation analysis; and (iv) its separate predictions for the EODs and scaled VDCs offer a novel approach to modeling functional degradation data residing on heterogeneous domains.
	
	The rest of the paper is as follows. In Section \ref{se:data}, we introduce the NASA battery dataset for the analysis. We discuss insights we gain from the existing work on the dataset as well as their gaps in modeling VDCs that inspire this work. In Section \ref{se:model}, we present the methodology in detail including notation, modeling, prediction, and degradation analysis. In Section \ref{se:simulation}, we conduct a simulation study to compare our method with a standard method. In Section \ref{se:analysis}, we present the real data analysis using the developed framework. Finally, we conclude with some summaries and areas for future research in Section~\ref{se:conclusion}.

	\section{Battery Life Study from NASA Ames Prognostics Data Repository}\label{se:data}
	
\begin{figure}
		\centering
		\includegraphics[width=0.55\linewidth]{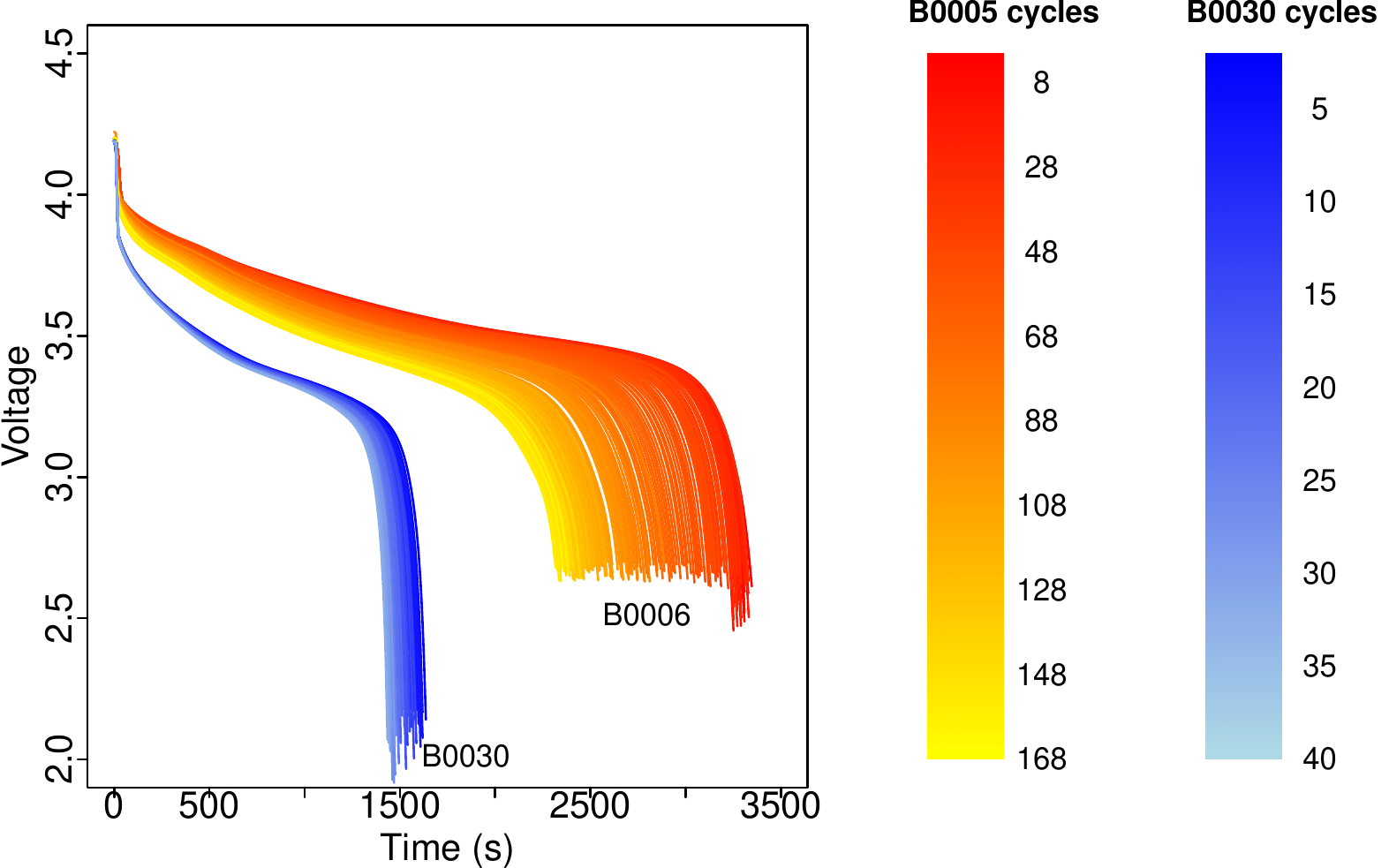}
		\caption{Voltage discharge curves by cycles number for battery B0005 (discharge current: 2 Amps, Stopping voltage: 2.7 Volts, temperature: $24^\circ$C) and battery B0030 (discharge current: 4 Amps, Stopping voltage: 2.2 Volts, temperature: $43^\circ$C).}
		\label{fig:discharge_curves}
	\end{figure}

	The study we consider here is from the NASA Prognostics Data Repository \citep{saha2007battery}. In the study, Li-ion batteries were tested through cycles of charge and discharge at different room temperatures and conditions. Experimenters performed the same charging procedure for all batteries after each discharge cycle. The lifetime measure for batteries was the number of discharge cycles. The experimental conditions for discharge cycles varied among batteries. We focus on the following conditions.
	\begin{itemize}
		\item Testing temperature (\emph{temp}): room temperature ($24\degreeC$), elevated temperature ($43\degreeC$), or low temperature ($4\degreeC$).
		\item Discharge current (\emph{dc}): 1 Amp, 2 Amps, or 4 Amps.
		\item Level of voltage where discharge ends or stopping voltage (\emph{sv}): 2 Volts, 2.2 Volts, 2.5 Volts, or 2.7 Volts.
	\end{itemize}
	The data were collected from 20 batteries with constant experimental conditions across discharge cycles in the experiment. Throughout the discharge cycles of each battery, measurements of discharge voltage levels were recorded on grids of unnecessarily equally-spaced time points. While the range of the number of grid points was 3 to 548, out of the 1985 total cycles, only 3 cycles had fewer than 100 grid points. Out of the remaining 1982 densely sampled cycles, 1942 cycles had 100 to 400 grid points. As shown in Figure \ref{fig:discharge_curves} for two batteries, the VDCs for each battery clearly show a degrading pattern. In early discharge cycles, the battery voltage stays high for a longer time until a sharp drop at a later time in the cycle. As the cycle number increases, the battery voltage sustains for a shorter time and drops sharply at an earlier time in the cycle. This pattern is revealed by the motion of the discharge curves moving inward in Figure \ref{fig:discharge_curves}.
	
	\begin{figure}
		\centering
            \includegraphics[width=0.8\linewidth]{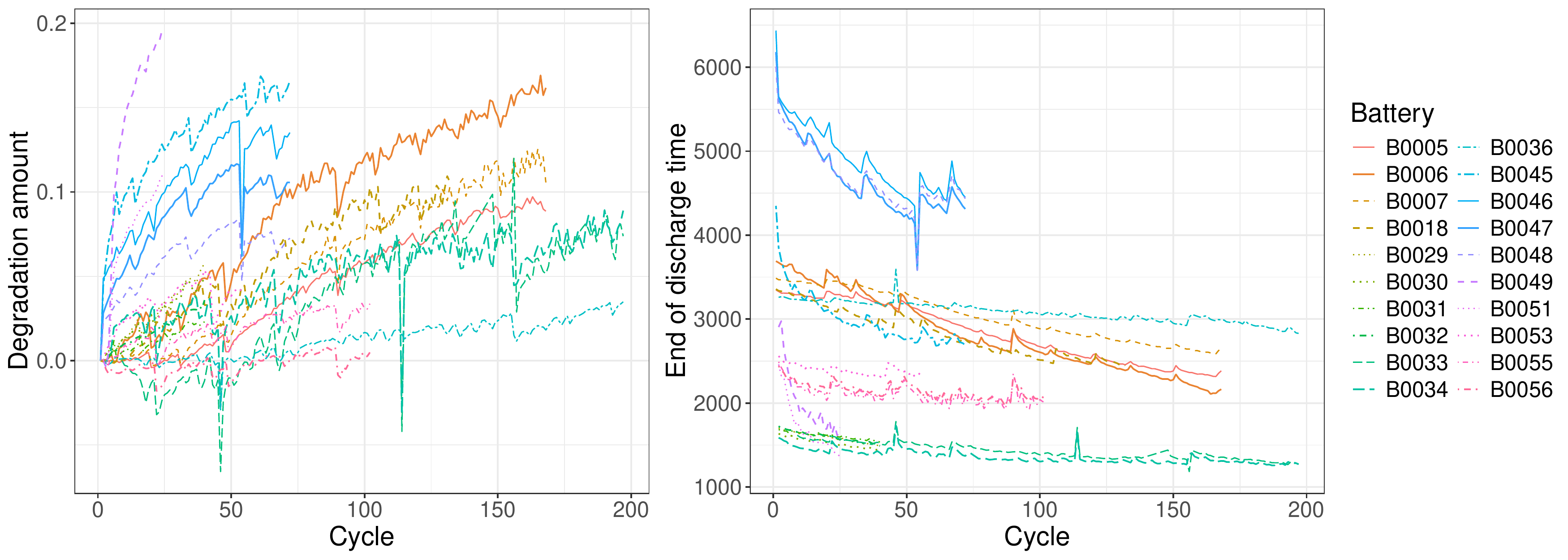}
		\caption{Left: degradation amount by cycle number and battery determined by $L^1$-norm of batteries' VDCs. Right: EOD by cycle number and battery.}
		\label{fig:degradation}
	\end{figure}
	
	This dataset is well-known in battery health prognostics, whose main focus is on studying the state of health (SOH) of batteries with the goal of predicting the remaining useful life (RUL). The RUL is determined by how close a battery's capacity is to a preset threshold. As a standard measurement, the battery capacity at a cycle is closely related to the area under the corresponding VDC. Therefore, most existing methods have focused on directly modeling and predicting battery capacity along discharge cycles on the degradation path.
	
	Our objective is to predict battery degradation in future discharge cycles for all 20 batteries taking into account their experimental conditions. Figure \ref{fig:degradation} depicts the degradation paths of all 20 batteries over recorded discharge cycles, where the degradation amount is defined in \eqref{eq:dic} based on the $L^1$-norms of VDCs. A majority of batteries have linear degradation pattern with some having slightly curved paths. Along the degradation paths, there are several jumps, adding to the complexity of the modeling. Some batteries even displayed negative degradation amounts due to their irregular VDCs that actually had slightly enlarged $L^1$-norms. There are useful insights we can gain from the work above that could aid in our prediction approach. For example, \citet{saha2009pf} mentioned the importance of considering the rest period between discharge cycles, which, denoted by $\Delta_{ic}$, is the elapsed time between cycles $c-1$ and $c$ of battery $i$. The effect of $\Delta_{ic}$ is known as self-recharge, a phenomenon where battery capacity increases after resting. It is due to the dissipation of build-up materials around electrodes. The authors suggest accounting for $\Delta_{ic}$ in the form of an exponential process. Since an increase in battery capacity leads to a larger EOD, we use $\exp(-1/\Delta_{ic})$ as a covariate in modeling the EOD. \citet{saha2009pf}, \citet{liu2012nlar}, and \citet{liu2014fusion} also considered autoregressive effects in their modeling of battery capacity and found that the inclusion of an order-one autoregressive term in the EOD model would significantly increase its prediction power.
	
	In the next section, we describe the details of our functional degradation models. Instead of relying only on historical capacity measurements, we take full advantage of all the information contained in voltage curves from the past discharge cycles to make predictions of the entire voltage discharge curves in future discharge cycles. Predicted VDCs allow battery users the access of specific features that can be extracted from a complete VDC. For example, besides the EOD of a discharge cycle, the decomposition of a VDC curve into components for an impedance model can be useful for discharge performance evaluation and voltage unloading within the cycle \citep{luo2011study}. Once voltage discharge curves are available, a degradation measure can be calculated based on the chosen health indicator, leading to more versatile battery health prognostics as demonstrated in Section \ref{se:analysis}.

	\section{Functional Degradation Model}\label{se:model}
	
	\subsection{Notation}\label{ss:notation}
	
	Suppose that there are $n$ batteries. For battery $i$, $n_i$ cycles are tested with each cycle producing a discharge curve. Let $y_{ic}(r)$ be the original discharge curve for battery $i$ at cycle $c$, $c=1,\cdots, n_i$, and $i=1, \cdots, n$. Here the discharge time is $r\in [0, b_{ic}]$, with $b_{ic}$ being the end time, or EOD, for this cycle. The covariate vector $\zvec_{ic}$ of length $m$ represents all the covariates at cycle $c$ for battery $i$, which also include the cycle-invariant experimental conditions such as the temperature, the discharge current, and the stopping voltage.
	
\begin{figure}
		\centering
		\includegraphics[width=0.6\linewidth]{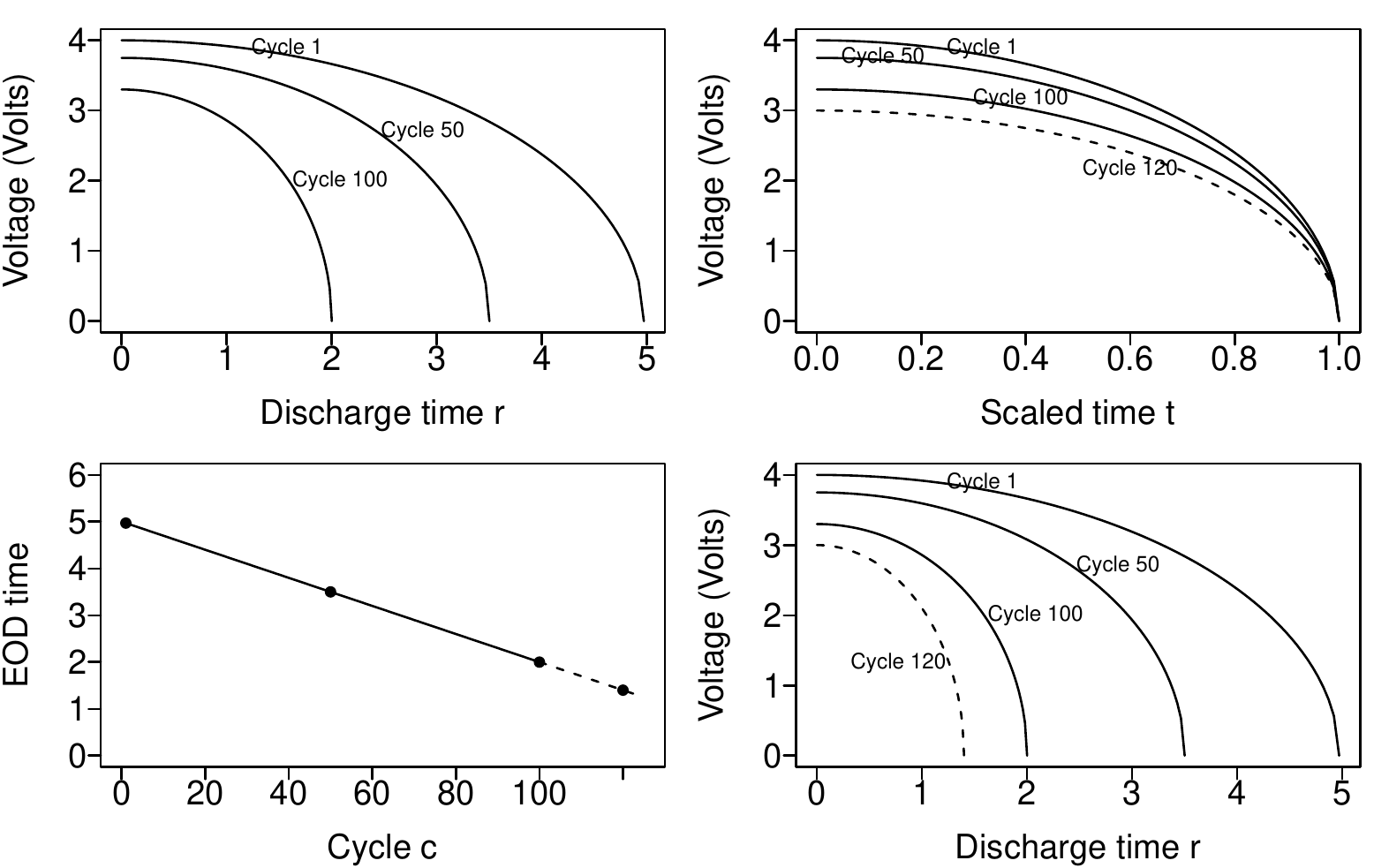}
		\caption{Diagram of voltage discharge curve processing for a toy example of battery. Top left: VDCs $y_{ic}(r)$ from three example cycles ($c=1, 50, 100$) plotted on their original discharge time domain (solid curves). Top right: scaled VDCs $x_{ic}(t)$ with discharge time rescaled by the EODs. Dashed line is the predicted curve for cycle 120. Bottom left: EODs $b_{ic}$ plotted against the cycle index $c$. Dashed part is the predicted EOD after cycle 100. Bottom right: Predicted VDC for cycle 120 superimposed on the plot of observed VDCs on the discharge time domain.}
		\label{fig:one_unit_diagram}
	\end{figure}

	To make the discharge curves comparable across different cycles, we first re-scale the discharge time by the EOD to obtain a scaled curve. Namely, we consider the re-scaled time $t=r/b_{ic}$ such that $t\in [0,1]$. And the scaled curve is $x_{ic}(t)=y_{ic}(r)=y_{ic}(b_{ic}t)$. Now each of the original discharge curves is represented by the pair $(b_{ic}, x_{ic}(\cdot))$. Then the observations for battery $i$ become $(b_{ic}, x_{ic}(\cdot),\zvec_{ic})$, a mix of scalar, functional and vector components. The first three plots in Figure \ref{fig:one_unit_diagram} provides an illustration of this data processing step.
	
	Our goal is to model $y_{ic}(r)|\zvec_{ic}$, that is, to model $(b_{ic}, x_{ic}(t))|\zvec_{ic}$, a response consisting of the scalar component $b_{ic}$ and the functional component $x_{ic}(t)$. Here we consider a two-step modeling of $x_{ic}(t)|\zvec_{ic}$ and then $b_{ic} | (x_{ic}(t), \zvec_{ic})$.
	
	\subsection{Step 1: Predictive modeling of $x_{ic}(t)$}\label{ss:step1}

	The first step is to develop a predictive model for the scaled curves $x_{ic}(\cdot)$ with the incorporation of the covariates $\zvec_{ic}$. The model must have the ability to predict the scaled curve $x_{ic}(\cdot)$ at a future cycle $c$. This is achieved through performing an FPCA on $x_{ic}(\cdot)$ and then develop a linear regression model with the FPC scores as the responses against the cycle index number $c$ and experimental conditions $\zvec_{ic}$.
	
	Suppose $\Sigma(s,t)$ is the covariance function for the random functions $x_{ic}(t)$. Let $\lambda_j$ and $\phi_j(t)$, $j=1,2,\ldots$, be the eigenvalues and eigenfunctions of $\Sigma(s,t)$. Then $x_{ic}(t)$ admits the Karhunen-Lo\`{e}ve expansion
	\begin{equation}\label{eq:kld}
		x_{ic}(t) = \mu(t)+\sum_{j=1}^\infty \gamma_{ic,j}\phi_j(t),
	\end{equation}
	where $\mu(t)$ is the mean function and $\gamma_{ic,j}=\int_0^1 x_{ic}(t)\phi_j(t)dt$ are uncorrelated random coefficients with mean 0 and variance $\lambda_j$. In practice, a truncation is often applied such that
	\begin{equation}\label{eq:decom}
		x_{ic}(t) \approx \mu(t)+\sum_{j=1}^K \gamma_{ic,j}\phi_j(t),
	\end{equation}
	where $K$ is a pre-determined truncation point, usually selected to achieve a certain percentage of total variation explained (e.g., 80\% or 90\%).
	Let $\boldsymbol{\gamma}_{ic}=(\gamma_{ic,1},\ldots,\gamma_{ic,K})^\top$. We consider the multivariate mixed-effects model for modeling the FPC scores of the scaled VDCs,
	\begin{equation}
		\label{eq:gamma}
		\boldsymbol{\gamma}_{ic}= \mathbf{v}_0 + \mathbf{u}_{0i}+(\mathbf{v}_1+\mathbf{u}_{1i})c+ \boldsymbol{P}\zvec_{ic} +
		\boldsymbol{\delta}_{ic},
	\end{equation}
	where $\mathbf{v}_0=(v_{01},\ldots,v_{0K})^\top$ is the vector of intercepts, $\mathbf{u}_{0i}=(u_{01i},\ldots,u_{0Ki})^\top$ is the vector of random intercepts for battery $i$, $\mathbf{v}_1=(v_{11},\ldots,v_{1K})^\top$ is the vector of slopes, $\mathbf{u}_{1i}=(u_{11i},\ldots,u_{1Ki})^\top$ is the vector of random slopes for battery $i$, $\boldsymbol{P} = (\mathbf{p}_{1},\dots,\mathbf{p}_{m})$ is a $K\times m$ matrix with column $h$ being $\mathbf{p}_h = (p_{h1},\dots,p_{hK})^\top$, the vector of slopes for covariate $h$, $\zvec_{ic}=(z_{1,ic},\dots,z_{m,ic})^\top$ is the vector of $m$ covariates for battery $i$ at cycle $c$, and $\boldsymbol{\delta}_{ic} = (\delta_{ic,1},\dots,\delta_{ic,K})^\top$ is the vector of random errors.
	
	For this step, we first perform the FPCA on all the scaled curves $x_{ic}(t)$, take the first $K$ FPCs $\{\hat{\phi}_j(t): j=1,\ldots, K\}$, and then compute the $K$ FPC scores $\{\hat{\gamma}_{ic,j},j=1,\ldots,K\}$ corresponding to each curve $x_{ic}(t)$. These FPC scores are then fed into the multivariate mixed-effects model \eqref{eq:gamma} as responses. The parameters to be estimated include the fixed-effects components, $\{\boldsymbol{v}_0, \boldsymbol{v}_1,\boldsymbol{P}\}$, and the distributional parameters of the random effects. Suppose $\boldsymbol{U} = (\mathbf{u}_{0i}^\top,\mathbf{u}_{1i}^\top)^\top \sim \mathcal{N}(\mathbf{0},\Sigma_{U})$ and $\boldsymbol{\delta}_{ic} \sim \mathcal{N}(\mathbf{0},\Sigma_{\delta})$. We can apply the standard tool for multivariate linear mixed-effects models (MLMMs) to estimate the fixed effects $\{\boldsymbol{v}_0, \boldsymbol{v}_1,\boldsymbol{P}\}$, and the covariance matrices $\Sigma_U$ and $\Sigma_{\delta}$ of the random effects. Let $\boldsymbol{\Gamma}_{i} = (\boldsymbol{\gamma}_{i\ldotp,1},\boldsymbol{\gamma}_{i\ldotp,2},\dots,\boldsymbol{\gamma}_{i\ldotp,K})$, where each $\boldsymbol{\gamma}_{i\ldotp,k}$ is the $n_i\times 1$ vector of the $k^{th}$ FPC scores for the $n_i$ cycles of battery $i$. Following the suggestions in \citet{shah1997random} and \citet{thiebaut2002bivariate}, we vectorize the response matrix into $\boldsymbol{y}_i = \mathbf{vec}(\Gamma_i) = (\boldsymbol{\gamma}_{i\ldotp,1}^\top,\boldsymbol{\gamma}_{i\ldotp,2}^\top,\dots,\boldsymbol{\gamma}_{i\ldotp,K}^\top)^\top$, and rewrite $\boldsymbol{\delta}_{ic}$ as $\boldsymbol{e}_i = (\boldsymbol{\delta}_{i\ldotp,1}^\top,\boldsymbol{\delta}_{i\ldotp,2}^\top,\dots,\boldsymbol{\delta}_{i\ldotp,K}^\top)^\top$ with $\boldsymbol{\delta}_{i\ldotp,j}=(\delta_{i1,j},\dots,\delta_{in_i,j})^\top$. Then the model becomes
	\begin{equation}\label{eq:mlmm_stacked}
		\boldsymbol{y}_i = \boldsymbol{X}^{*}_i\vvec + \boldsymbol{Z}^{*}_i\uvec_i + \boldsymbol{e}_i.
	\end{equation}
	Here $\boldsymbol{X}^{*}_i$ is the ${Kn_i\times q^{*}}$ design matrix collecting all the fixed-effects predictors such as cycle number, experimental covariates $\zvec_{ic}$, and indicator vector representing the membership, among the $K$ components, of the FPC score in the response. Matrix $\boldsymbol{Z}^{*}_i$ with dimension ($Kn_i)\times r^{*}$ is the random design matrix. Vector $\vvec$ is the vectorized version of the fixed effect parameters $\{\boldsymbol{v}_0, \boldsymbol{v}_1,\boldsymbol{P}\}$, while $\uvec_i$ collects individual random effects. We then estimate model \eqref{eq:mlmm_stacked} via the maximum likelihood or restricted maximum likelihood (REML) approach.
	
	\subsection{Step 2: Modeling $b_{ic} | x_{ic}(t)$ and its parameter estimation}\label{ss:step2}
	
	The second step is to develop a model of the EOD $b_{ic}$ given the scaled curve $x_{ic}(t)$. The observed EODs and scaled curves can be used to obtain the estimates of the parameters in the following models. Then the fitted model can be used to predict a future EOD given the predicted scaled curve from Step 1. Note that the EOD $b_{ic}$ is a scalar measurement with a longitudinal nature. We consider two versions of mixed-effects models, where experimental conditions are part of the fixed effects. In addition, the second model also includes the VDC $x_{ic}(\cdot)$ as a functional covariate. That is,
	\begin{equation}\label{eq:bic_lme}
	 b_{ic} = \alpha_0+w_{0i} + (\alphavec_1+\wvec_{1i})^\top \zvec_{1,ic} + \alphavec_2^\top \zvec_{2,ic} +\epsilon_{ic},
	\end{equation}
	\begin{equation}\label{eq:bic_flmm}
	 b_{ic} = \alpha_0+w_{0i} + (\alphavec_1+\wvec_{1i})^\top \zvec_{1,ic} + \alphavec_2^\top \zvec_{2,ic} + \int_0^1 \left(\beta(t) + b_i(t)\right)x_{ic}(t)dt  +\epsilon_{ic},
	\end{equation}
	where 
 $\zvec_{1,ic}=(z_{11,ic},\cdots,z_{1m_1,ic})^\top$ are covariates with both fixed and random effects, $\alpha_0$ and $w_{0i}$ are respectively the fixed and random intercepts, $\alphavec_1=(\alpha_{11},\cdots,\alpha_{1m_1})^\top$   and $\wvec_{1i}=(w_{11,i},\cdots,w_{1m_1,i})^\top$ are respectively the fixed and random slopes, $\alphavec_2=(\alpha_{21},\cdots,\alpha_{2m_2})^\top$ is the coefficient vector for $m_2$ fixed-effect-only covariates $\zvec_{2,ic}=(z_{21,ic},\cdots,z_{2m_2,ic})^\top$, $\epsilon_{ic}$ are i.i.d. random errors with mean 0 and variance $\sigma^2_{\epsilon}$, and $\beta(t)$ and $b_i(t)$ are respectively the unknown fixed and random coefficient functions modeling the effects of the scaled voltage discharge curves $x_{ic}(t)$. The random effects, $\wvec_i=(w_{0i},\wvec_{1i}^\top)^\top$ are assumed to follow $\mathcal{N}(0,\sigma^2_{\epsilon}\Psivec)$, where $\Psivec$ is the $(m_1+1)\times(m_1+1)$ covariance matrix of $\wvec_{i}$ scaled by $\sigma^2_{\epsilon}$. Note that although we use the same notation $\zvec_{1,ic}$ and $\zvec_{2,ic}$ for covariates in models \eqref{eq:bic_lme} and \eqref{eq:bic_flmm}, they don't have to use the same set of covariates. For example, some covariates already in the model for $x_{ic}(t)$ (or $\boldsymbol{\gamma}_{ic}$) at Step 1 may not necessarily be included in model \eqref{eq:bic_flmm}. 
 
  Model \eqref{eq:bic_lme} is a standard linear mixed-effects model, and can be estimated through standard software procedures for linear mixed-effects models. 
	Model \eqref{eq:bic_flmm} is a type of functional linear mixed-effects (FLMM) model. \citet{liu2017estimating} suggested an expectation-maximization (EM) REML-based algorithm to fit a functional linear model that allows mixed-effects for both scalar and functional covariates. To fit this model, we can expand $\beta(t)$ and $b_{i}(t)$ as linear combinations of basis functions,
	e.g., B-spline basis functions. Suppose the expansions are respectively
	$\beta(t) = \sum_{j=1}^Rp_{j}\phi_j(t)=\phivec^{\top}(t)\mathbf{p}$ where $R$ is the number of basis functions chosen to represent $\beta(t)$
	and
	$b_{i}(t) = \sum_{k=1}^Sq_{ik}\psi_k(t)=\psivec^{\top}(t)\mathbf{q}_i$ where $S$ is the number of basis functions chosen to represent each $b_{i}(t)$.
	Since $b_{i}(t)$ is a zero-mean Gaussian process, we can assume the coefficient vector $\mathbf{q}_i \sim \mathcal{N}(0,\sigma^2_{\epsilon}\mathbf{D})$ for some covariance matrix $\mathbf{D}$ such that $\gamma(s,t)=\sigma^2_\epsilon\psivec^\top(s)\mathbf{D}\psivec(t)$.
	Upon substituting the expansions of $\beta(t)$ and $b_i(t)$, we can rewrite \eqref{eq:bic_flmm} as
	$$
	b_{ic} = \mathbf{g}_{ic}^\top\alphavec + \mathbf{h}_{ic}^\top\mathbf{w}_i + \mathbf{j}_{ic}^\top\mathbf{p} + \mathbf{k}_{ic}^\top\mathbf{q}_i + \epsilon_{ic},
	$$
	where $\mathbf{g}_{ic} = (1,\zvec_{1,ic}^\top, \zvec_{2,ic}^\top)^\top$ and $\mathbf{h}_{ic}=(1,\zvec_{1,ic}^\top)^\top$ are respectively the design matrices of the fixed and random components, $\mathbf{j}_{ic} = \int_0^1 x_{ic}(t)\phivec(t)dt$ and $\mathbf{k}_{ic} = \int_0^1 x_{ic}(t)\psivec(t)dt$.
	Then the task becomes estimating the basis expansion coefficients $\mathbf{p}=(p_1,\dots,p_R)^\top$ and $\mathbf{q}_i = (q_{i1},\dots,q_{iS})^\top$, the intercept and slopes $\alphavec=(\alpha_0,\alphavec_1^\top,\alphavec_2^\top)^\top$ in the fixed effect, and the variance components $\sigma^2_\epsilon$, $\Psivec$ and $\mathbf{D}$. Given $\sigma^2_\epsilon$, $\Psivec$ and $\mathbf{D}$, the objective function for $\thetavec = (\alphavec^\top, \mathbf{p}^\top)^\top$ and $\varthetavec_i = (\wvec_i^\top, \mathbf{q}_i^\top)^\top$ is
	
	\begin{multline}\label{eq:obj_element}
			M\left(\thetavec, \{\varthetavec_i\}_{i=1}^n,\sigma^2_\epsilon,\mathbf{D}\right) = \sum_{i=1}^n\sum_{c=1}^{n_i}\frac{1}{2\sigma^2_\epsilon}\left( b_{ic} - \mathbf{g}_{ic}^\top\alphavec - \mathbf{h}_{ic}^\top\mathbf{w}_i - \mathbf{j}_{ic}^\top\mathbf{p} - \mathbf{k}_{ic}^\top\mathbf{q}_i \right)^2 \\
			+ \frac{\lambda_\beta}{2}\int_0^1 \{\beta''(t)\}^2dt + \frac{\lambda_b}{2}\sum_{i=1}^n\int_0^1 \{b_i''(t)\}^2dt 
			+  \frac{1}{2\sigma^2_\epsilon}\sum_{i=1}^n\mathbf{q}_i^\top \mathbf{D}^{-1} \mathbf{q}_i + \frac{1}{2\sigma^2_\epsilon}\sum_{i=1}^n\wvec_i^\top\Psivec^{-1}\wvec_i.
	\end{multline}
	It consists of a least squares component, the roughness penalties for $\beta(t)$ and $b_i(t)$ with $\lambda_\beta$ and $\lambda_b$ be the corresponding smoothing parameters, and the REML components for the random components $\mathbf{w}_i$ and $\mathbf{q}_i$. 
 Let $\mathbf{b}_i = (b_{i1},\dots,b_{in_i})^\top$, $\tilde{\mathbf{G}}_i = (\tilde{\mathbf{g}}_{i1}, \dots, \tilde{\mathbf{g}}_{in_i})^\top$ with $\tilde{\mathbf{g}}_{ic} = (\mathbf{g}_{ic}^\top,\mathbf{j}_{ic}^\top)^\top$, and $\tilde{\mathbf{H}}_i = (\tilde{\mathbf{h}}_{i1}, \dots, \tilde{\mathbf{h}}_{in_i})^\top$ with $\tilde{\mathbf{h}}_{ic} = (\mathbf{h}_{ic}^\top,\mathbf{k}_{ic}^\top)^\top$. Define the matrices $G_{\phi} = \int^1_0 \frac{d^2\phi(t)}{dt^2}\left(\frac{d^2\phi(t)}{dt^2}\right)^\top dt$ and $G_{\psi} = \int^1_0 \frac{d^2\psi(t)}{dt^2}\left(\frac{d^2\psi(t)}{dt^2}\right)^\top dt$. Then a full matrix-vector form of the objective function $\eqref{eq:obj_element}$ is
	
	\begin{multline}\label{eq:obj_matrix}
			M\left(\thetavec, \{\varthetavec_i\}_{i=1}^n,\sigma^2_\epsilon,\mathbf{D}\right) = \sum_{i=1}^n\frac{1}{2\sigma^2_\epsilon}\parallel \mathbf{b}_{i} - \tilde{\mathbf{G}}_{i}\thetavec - \tilde{\mathbf{H}}_{i}\varthetavec_i\parallel^2 \\+ \frac{\lambda_\beta}{2}\mathbf{m}^\top G_{\phi} \mathbf{m} + \frac{\lambda_b}{2}\sum_{i=1}^n \mathbf{q}_i^\top G_{\psi} \mathbf{q}_i 
			+  \frac{1}{2\sigma^2_\epsilon}\sum_{i=1}^n\mathbf{q}_i^\top \mathbf{D}^{-1} \mathbf{q}_i + \frac{1}{2\sigma^2_\epsilon}\sum_{i=1}^n\wvec_i^\top\Psivec^{-1}\wvec_i.
	\end{multline}
The objective function \eqref{eq:obj_matrix} can be minimized with closed-form solutions for $\thetavec$ and $\varthetavec_i$. To estimate $\sigma^2_\epsilon$ and $\mathbf{D}$, we can employ an iterative REML-based EM algorithm starting with some initial values for $\sigma^{2(0)}_\epsilon$ and $\mathbf{D}^{(0)}$. The M-step computes the estimates of $\thetavec^{(r)}$ and $\varthetavec_i^{(r)}$ as the minimizer of \eqref{eq:obj_matrix}, followed by the E-step of updating $\sigma^{2(r)}_\epsilon$ and $\mathbf{D}^{(r)}$. The procedure is repeated until all the estimators converge within a pre-specified tolerance level.
	We employ multi-fold cross validation for both the selection of smoothing parameters $\lambda_\beta$ and $\lambda_b$ and the selection of models and covariates to include in $\{\zvec_{1,ic},\zvec_{2,ic}\}$.

	\subsection{Step 3: Estimation, prediction, and degradation analysis of $y_{ic}(r)$}\label{ss:step3}
	
	Once the scaled discharge curve $x_{ic}(t), t\in [0,1]$, and the EOD $b_{ic}$ are estimated through the two steps described above, the estimate for the standard discharge curve $y_{ic}(r)$ on its natural domain, $r\in [0,b_{ic}]$, can be constructed point-wisely by simply re-scaling the estimate $\hat{x}_{ic}(t)$ with the EOD estimate $\hat{b}_{ic}$. The prediction of the voltage discharge curve $\hat{y}_{i\tilde{c}}(r)$ for a future cycle $\tilde{c}$ can be obtained similarly through the predictions $\hat{x}_{i\tilde{c}}(t)$ and $\hat{b}_{i\tilde{c}}$. Then $\hat{y}_{i\tilde{c}}(r)$ is used as a functional measurement prediction for the following degradation analysis.
	
\begin{figure}
		\centering
		\includegraphics[width=0.6\linewidth]{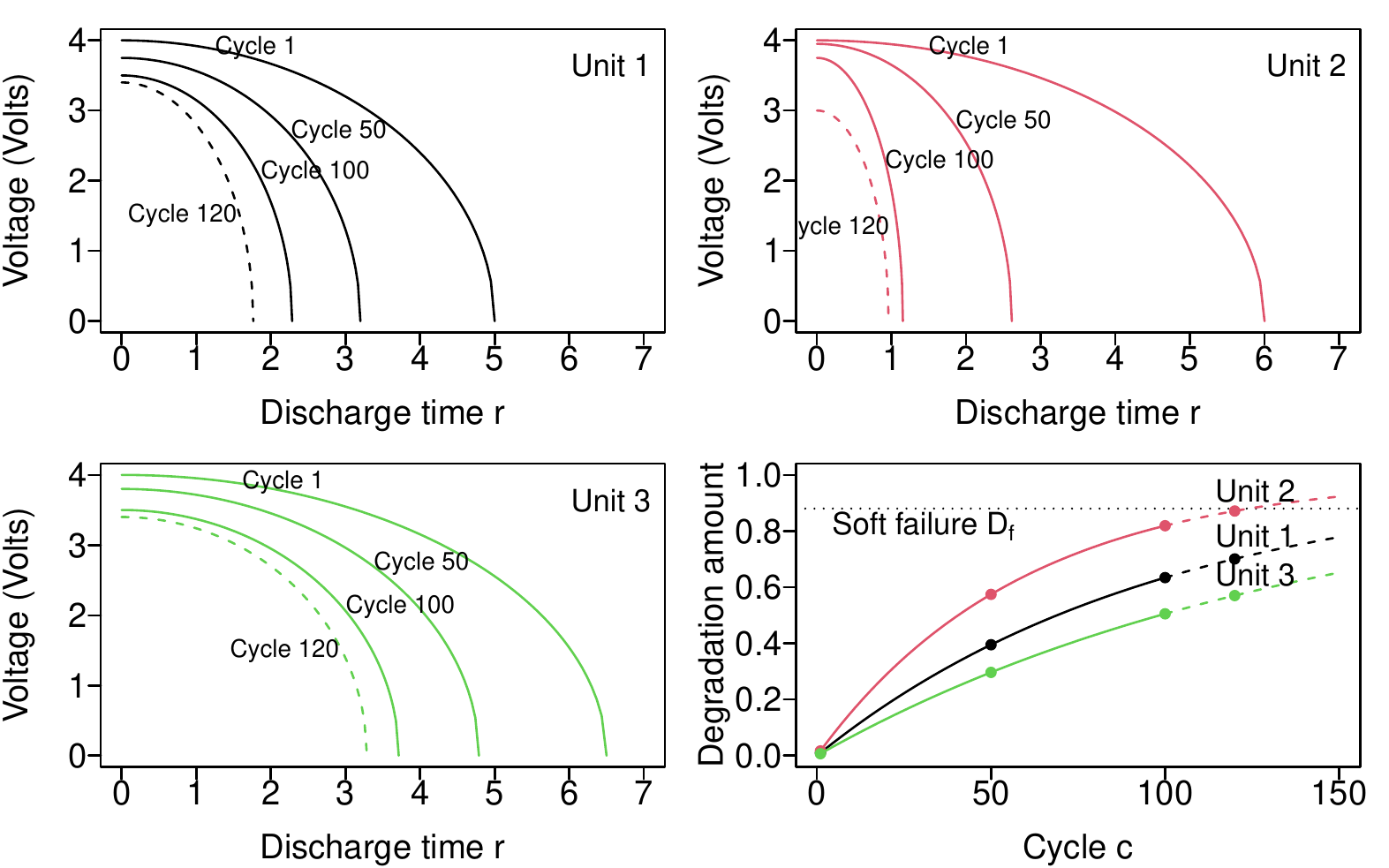}
		\caption{Degradation analysis diagram for multiple batteries. Panel 1 (top left), Panel 2 (top right), and Panel 3 (bottom left): Voltage discharge curves of each unit (battery). Solid curves are for modeling under procedure proposed in Section \ref{se:model}. Dashed curves are predictions based on final model. For each of these VDCs, degradation amount is measured according to equation \eqref{eq:dic}. Panel 4 (bottom right): degradation paths for all units obtained from VDCs with observed (solid) and predicted (dashed) values overlayed with a soft failure threshold $D_f$.}
		\label{fig:multiple_units_diagram}
	\end{figure}

	One advantage of producing predicted VDCs in the original functional form is the flexibility in performing degradation analysis according to the degradation definition set by practitioners. One can examine the full predicted VDCs and anticipate discharge voltage behavior within a future discharge cycle. We illustrate this approach in more detail in Section \ref{ss:da_fulldata}. Another popular approach is to convert a full VDC into a scalar degradation measurement. As demonstrated in Section \ref{se:simulation}, existing methods construct degradation path models directly from a scalar degradation measurement $\Upsilon_{ic}$ calculated from the observed VDC $y_{ic}(\cdot)$. Such a scalar measurement could be the $L^p$-norm of the $y_{ic}(\cdot)$. That is,
	\begin{equation}\label{eq:lp}
		\Upsilon_{ic} = ||y_{ic}(\cdot)||_p \equiv \left(\int_0^{b_{ic}}|y_{ic}(r)|^pdr\right)^{1/p}.
	\end{equation}
	When $p=1$, $\Upsilon_{ic}$ is the area under the discharge curve, which is directly related to battery capacity. To compare a battery's degradation at cycle $c$ relative to its original state, i.e., the first discharge cycle, we can further compute the degradation amount
	\begin{equation}\label{eq:dic}
		d_{ic} = \frac{\Upsilon_{i1}-\Upsilon_{ic}}{\Upsilon_{i1}} = \frac{|| y_{i1}(\cdot) ||_p - || y_{ic}(\cdot) ||_p}{|| y_{i1}(\cdot) ||_p}.
	\end{equation}
	The degradation amount $d_{ic}$ in \eqref{eq:dic} can be interpreted as the difference between the $L^p$-norm of the discharge curve at the first cycle and the $L^p$-norm of the discharge curve at cycle $c$ for battery $i$. This measure can be compared to any soft failure threshold $D_f$ set by practitioners to represent the end of rechargeable battery usability. Figure \ref{fig:multiple_units_diagram} demonstrates the degradation analysis procedure on a toy example. We consider three hypothetical batteries or units (Panels 1 to 3). Each unit starts with some observed VDCs (solid curves), which yield a predicted VDC (dashed curve) based on a prediction model. We then compute a degradation amount for each VDC (observed or predicted). Then the degradation amounts for each battery are assembled to produce the degradation path in Panel 4. At last, the degradation path can be compared with the soft failure threshold to determine the status of the unit.

 Algorithm \ref{algorithm:proc} presents the complete algorithm for the proposed method.
 
   \begin{algorithm}
\DontPrintSemicolon
  
  \KwData{Raw curves $y_{ic}(\cdot)$ and covariates $\zvec_{ic}$ (including $\zvec_{1,ic}$ and $\zvec_{2,ic}$) for cycles $c=1,\ldots,n_i$ and units $i=1,\ldots,n$.}
  \textbf{Estimation Procedure:}\\
  \begin{itemize}
  \item \textbf{Data Transformation:} Compute EODs $b_{ic}$ and scaled curves $x_{ic}(\cdot)$ from raw curves $y_{ic}(\cdot)$. 
  \item \textbf{Modeling Scaled Curves $x_{ic}(\cdot)$:} Perform FPCA on all the scaled curves $\left\{\{x_{ic}(\cdot)\}_{c=1}^{n_i} \right\}_{i=1}^n$ to obtain FPC scores $\gammavec_{ic}$ and regress $\gammavec_{ic}$ against covariates $\zvec_{ic}$ using the multivaraite mixed-effects model \eqref{eq:gamma}.
  \item \textbf{Modeling EODs $b_{ic}$:} Fit the LME \eqref{eq:bic_lme} using the covariates $\zvec_{1,ic}$ and $\zvec_{2,ic}$, or the FLMM \eqref{eq:bic_flmm} using the functional covariate $x_{ic}(\cdot)$ in addition to the covariates $\zvec_{1,ic}$ and $\zvec_{2,ic}$.
  \end{itemize}
  \vspace{-0.2cm}
\textbf{Estimation, Prediction, and Degradation Analysis:} For cycle $c$ of battery $i$,
\begin{itemize}
\vspace{-0.2cm}
    \item \textbf{Estimation/Prediction:} compute the estimate/prediction $\hat{x}_{ic}(\cdot)$ of the scaled curve using \eqref{eq:gamma}; compute the estimate/prediction $\hat{b}_{ic}$ of 
    the EOD using \eqref{eq:bic_lme} or \eqref{eq:bic_flmm}; combine $\hat{x}_{ic}(\cdot)$ and $\hat{b}_{ic}$ to obtain the estimate/prediction $\hat{y}_{ic}(\cdot)$ of the original curve.
    \item \textbf{Degradation Analysis:}  compute the estimate/prediction $\hat{d}_{ic}$ of the degradation amount.
\end{itemize}
\caption{Functional Degradation Model\label{algorithm:proc}}
\end{algorithm}

	\section{Simulation Study}\label{se:simulation}

	\subsection{Simulation settings}
	
	We conduct a numerical simulation study to compare degradation analysis performance between our functional data approach and existing methods. In numerical studies, we simulate data from a hypothetical experiment with $n$ units. For each $i=1,\ldots,n$, unit $i$ is set to go through $n_i$ cycles of action that result in its degradation. Within each cycle, a curve observation is simulated. This curve is assembled from two components: a curve on the standard $[0,1]$ domain and a domain end point that scales the standard curve to an individualized domain. In particular, the domain end point $b_{ic}$ for curve $c$ of unit $i$ is simulated from the following versions of models \eqref{eq:bic_lme} or \eqref{eq:bic_flmm}:
	\begin{equation}\label{eq:bic_sim}
		b_{ic} = \alpha_0 + w_{0i} + (\alpha_1 + w_{1i})c  +  \alpha_2 b_{i,c-1} + \alpha_3 \exp\left\{\frac{-1}{\Delta_{ic}}\right\}+\alpha_4z_i+\epsilon_{ic},
	\end{equation}
	\begin{equation}\label{eq:bic_sim_flmm}
		b_{ic} = \alpha_0 + w_{0i} + (\alpha_2 + w_{2i})b_{i,c-1} + \alpha_3 \exp\left\{\frac{-1}{\Delta_{ic}}\right\}+ \int_0^1 \left(\beta(t) + b_i(t)\right)x_{ic}(t)dt+\epsilon_{ic},
	\end{equation} 
	where $w_{0i} \sim \N(0,\sigma^2_0)$, $w_{1i} \sim \N(0,\sigma^2_1)$, $w_{2i} \sim \N(0,\sigma^2_2)$, $\epsilon_{ic} \sim \N(0,\sigma^2_\epsilon)$. For the first cycle ($c=1$) we set $b_{i,0}=0$. The covariate $z_i$ represents a testing condition for unit $i$, which is generated from a uniform distribution on $[0,1]$. The time lag $\Delta_{ic}$ between cycles $c-1$ and $c$ is set up as follows. For the first cycle $\Delta_{i1}=0$. When $c$ is divisible by 10, we set $\Delta_{ic} = 10 + \ell$ where $\ell \sim \N(0,2^2)$. Otherwise, $\Delta_{ic} = 1 + \ell$, where $\ell \sim \N(0,0.1^2)$. In this manner, we assume the hypothetical system has a long break every 10 cycles.
	No matter which simulation settings are used, \eqref{eq:bic_sim} or \eqref{eq:bic_sim_flmm}, once $b_{ic}$ is generated, the curve $y_{ic}(r)$ with $r\in [0,b_{ic}]$ is obtained through $y_{ic}(r)=x_{ic}(r/b_{ic})$, where
 $x_{ic}(\cdot) = \mu(\cdot) +\sum_{j}^3\gamma_{ic,j}\phi_{j}(\cdot)$, $\mu(t)=0.75\log(60-59.5t)$, $\phi_{1}(t)=1$, $\phi_{2}(t)=\sqrt{2}\sin(2\pi t)$, and $\phi_{3}(t)=\sqrt{2}\cos(2\pi t)$ for $t\in[0,1]$.  
 Models for the coefficients $\gamma_{ic,j}, j = 1,2,3$, are
	\begin{equation}\label{eq:gammaicj_sim}
		\gamma_{ic,j} = (v_{0j} + u_{0ji}) + (v_{1j} + u_{1ji})c + v_{2}z_i + \delta_{ic,j},
	\end{equation}
	where $u_{0ji},u_{1ji}\sim \N(0,\sigma^2_u)$ and $\delta_{ic,j}\sim \N(0,\sigma^2_\delta)$.
	
\begin{figure}
		\centering
		\includegraphics[width=0.5\linewidth]{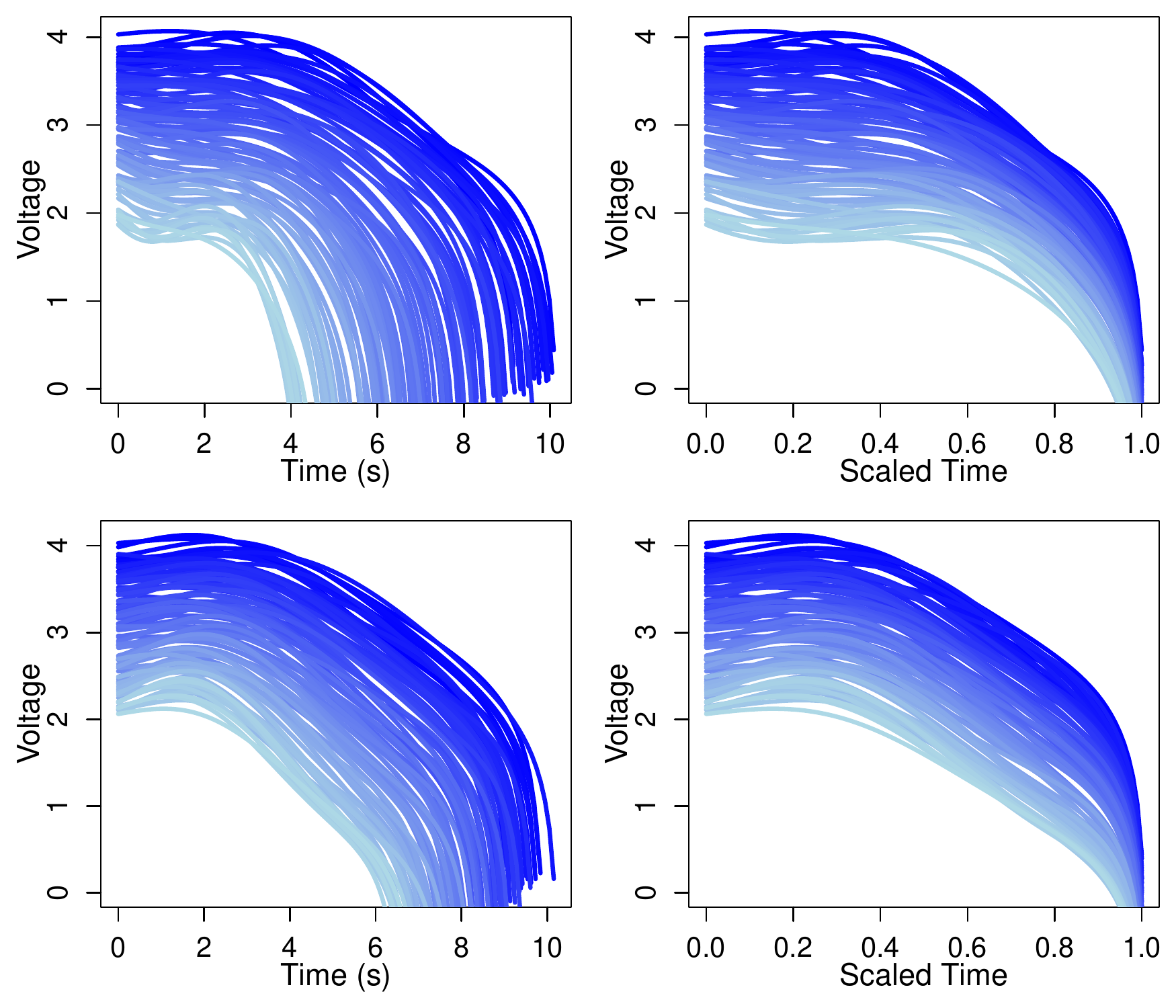}
  		\includegraphics[width=0.14\linewidth]{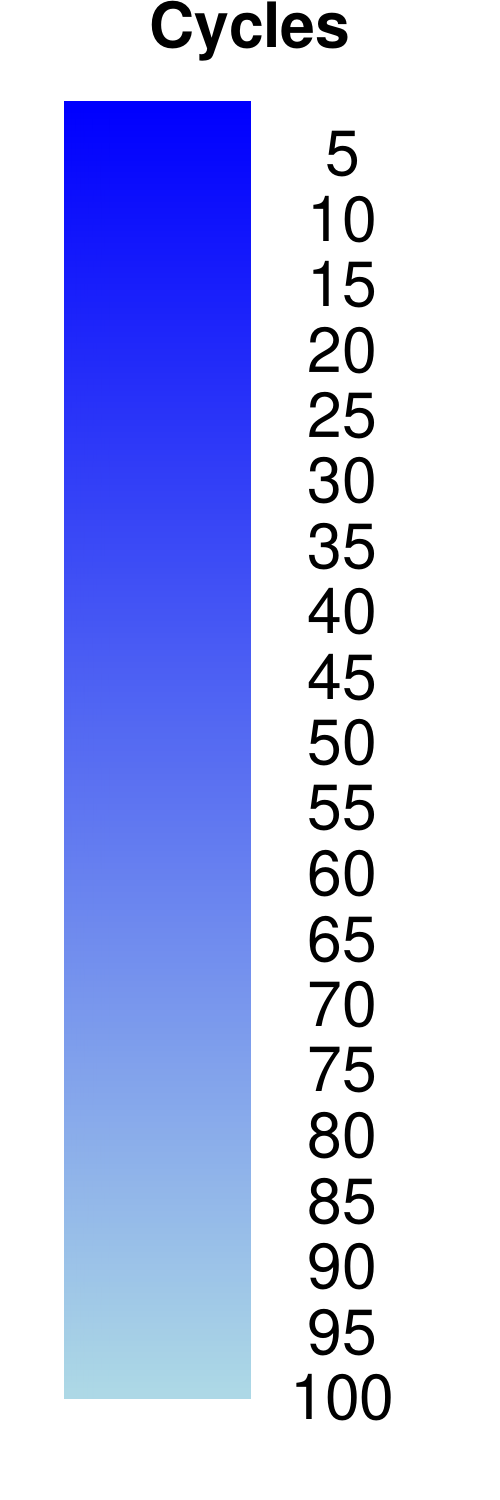}
		\caption{Example of simulated discharge curves through 100 discharge cycles for one unit. Top panels: simulation model \eqref{eq:bic_sim}; Bottom panels: simulation model \eqref{eq:bic_sim_flmm}. Left panels display the curves in their original domain. Right panels are the scaled curves.}
		\label{fig:simulated}
	\end{figure}

 With the guidance of the application, we choose our parameter values as follows. The fixed-effects coefficients in \eqref{eq:bic_sim} and \eqref{eq:bic_sim_flmm} are $\alpha_0=9$ for \eqref{eq:bic_sim}, $\alpha_0=3$ for \eqref{eq:bic_sim_flmm}, $\alpha_1=-0.06$, $\alpha_2=0.05$, $\alpha_3=1$, $\alpha_4=1$, and for $t \in [0,1]$, $\beta(t)=3-4t+\sin(\pi t)$ and $b_i(t)=r_{i1}+r_{i2}\sin(2\pi t)+r_{i3}\cos(2\pi t)$ with $\{r_{ij}\}_{j=1}^3$ following $\N(0,0.05^2)$, which are similar to functional slopes used in \citet{liu2017estimating}. The variances of random effects and noise are $\sigma_{0} = 0.9$ for \eqref{eq:bic_sim}, $\sigma_{0} = 0.3$ for \eqref{eq:bic_sim_flmm}, $\sigma_{1}=0.006$, $\sigma_{2}=0.005$, and $\sigma_{\epsilon} = 0.1$. In \eqref{eq:gammaicj_sim}, the fixed effects are set as $(v_{01},v_{02},v_{03})^\top = (1,0.1,-0.1)^\top$, $(v_{11},v_{12},v_{13})^\top = (-0.02,0,0)^\top$, and $(v_{21},v_{22},v_{23})^\top = (0.025,0.02,0.015)^\top$. The variances of random effects and noise are respectively $\sigma_u=0.001$ and $\sigma_{\delta}=0.05$. Figure \ref{fig:simulated} displays the data for a unit generated in this manner.
	
	To test the efficacy of the proposed method on the simulated data, we build predictive models from the first $tr\%$ of the number of cycles of each unit. $tr\%$ is the chosen training data ratio. Predictions are done on the remaining $1 - tr\%$. We choose the degradation measure of interest to be the normalized difference between the $L^1$-norms, or areas under the curves, of the VDCs at cycle $c$ and the first cycle, namely, $d_{ic} = (||y_{i1}(\cdot)||_1 - || y_{ic}(\cdot)||_1)/|| y_{i1}(\cdot)||_1$. The accuracy in prediction is evaluated by the root mean squared prediction error $\textrm{RMSPE} = \sqrt{\{\sum_{i=1}^n(n_i-\tilde{n}_i+1)\}^{-1}{\sum_{i=1}^n\sum_{c=\tilde{n}_i}^{n_i}(\hat{d}_{ic} - d_{ic})^2}}$, where $\tilde{n}_i = n_i\cdot tr\%+1$. We also evaluate the goodness of fit in estimating degradation amount using root mean squared error $\textrm{RMSE} = \sqrt{\{\sum_{i=1}^n(\tilde{n}_i-1)\}^{-1}{\sum_{i=1}^n\sum_{c=1}^{\tilde{n}_i - 1}(\hat{d}_{ic} - d_{ic})^2}}$. RMSPE and RMSE for scaled VDC and EOD time are defined in the supplementary section \ref{se:simul_supp}, which are similar to those for degradation.
	
    Our simulation settings have all the possible combinations of $n=\{20, 50, 100\}$, $n_i=\{50,100,150\}$, train ratio $tr\%=\{50\%, 80\%\}$, and EODs generated by \eqref{eq:bic_sim} or \eqref{eq:bic_sim_flmm}. Each simulation setting contains 100 data replications. The general path model is fitted with the \texttt{lme} routine from the R package \texttt{nlme}. For the simulation study, we apply the linear degradation path in equation \eqref{eq:linear_degrade}. For the proposed functional model approach, we apply the functional principal component decomposition from \texttt{fdapace} R package \citep{fdapace}, then the multivariate linear mixed-effects model for FPC scores are constructed using \texttt{lme} on the vectorized form as demonstrated above. Two models of estimating EOD time are considered. The first model is the linear mixed-effects model \eqref{eq:bic_sim} and the second model is the FLMM \eqref{eq:bic_sim_flmm}. We customized the code provided in \citet{liu2017estimating} to obtain the estimated parameters in the functional linear mixed model \eqref{eq:bic_sim_flmm}.

	\subsection{Existing degradation models}\label{ss:existing}
	Current methods in degradation analysis of rechargeable batteries do not take into account the full functional form of VDCs. The observed VDCs are transformed into a scalar degradation measurement through equations \eqref{eq:lp} and \eqref{eq:dic}. Then a general path model \citep{meeker2004degradation} is constructed based on the observed degradation amount
	\begin{equation}\label{eq:gpm}
		d_{ic} = \mathcal{D}_i(c;\betavec, \boldsymbol{\xi}_i|\zvec_{ic}) + \epsilon_{ic},
	\end{equation}
	where $\zvec_{ic}$ is the vector of covariates for battery $i$ at cycle $c$ and $\mathcal{D}_i$ is a function of the cycle $c$ given $\zvec_{ic}$ with parameters $\betavec$ and $\boldsymbol{\xi}$.
	In particular, $\betavec$ contains all the fixed parameters common across all batteries and $\boldsymbol{\xi}_i = (\xi_{i1},\dots,\xi_{i\ell})^\top$ collects $\ell$ random parameters representing battery-to-battery variations. In model \eqref{eq:gpm}, $\boldsymbol{\xi}_i$ is assumed to follow a multivariate normal distribution with mean $\mathbf{0}$ and covariance matrix $\boldsymbol{\Sigma}$. In this paper we consider $\ell=1$. The random errors $\epsilon_{ic}$ are assumed to be independent and identically distributed, following a normal distribution with mean 0 and variance $\sigma^2_\epsilon$. 
	
	The first step in the degradation model \eqref{eq:gpm} is specifying the functional form of $\mathcal{D}_{i}(c)$ based on the shape of the degradation path formed by the degradation amounts $d_{ic}$ along the cycles. Based on Figure \ref{fig:degradation}, we use a linear degradation path of the form
	\begin{equation}\label{eq:linear_degrade}
		\mathcal{D}_i(c) = \beta_0 + (\beta_1 + \xi_{i1})c + \zvec_{ic}^\top\mathbf{\beta}_z.
	\end{equation}
	By setting $\exp(\zeta_{i1})=|1+\xi_{i1}/\beta_1|$, we can see that model \eqref{eq:linear_degrade} has an equivalent form
	\begin{equation}\label{eq:linear_degrade_lognormal}
		\mathcal{D}_i(c) = \beta_0 + \beta_1\exp(\zeta_{i1})c + \zvec_{ic}^\top\mathbf{\beta}_z,
	\end{equation}
	where $\beta_1$ may differ from $\beta_1$ in \eqref{eq:linear_degrade} by a sign and the random effect $\zeta_{i1}$ now assumes a lognormal probability distribution.
	The goal is now to estimate parameters $\boldsymbol{\theta} =
	\{\boldsymbol{\beta}$, $\sigma^2_{\epsilon}$, $\boldsymbol{\Sigma}\}$.  Note that in general model \eqref{eq:gpm} is a nonlinear mixed-effects model, which can be estimated through standard methods such as maximum likelihood (ML) or REML. In practice, we apply the \texttt{nlme} and \texttt{lme} functions from the R package \texttt{nlme} \citep{pinheiro2017package}. 
 
 In the simulations, we use GPM to compare with the two versions of our functional degradation models (FDMs): the FDM-LME which is the FDM using the LME model \eqref{eq:bic_sim} for EODs, and the FDM-FLMM which is the FDM using the FLMM \eqref{eq:bic_sim_flmm}. GPM uses the same covariates as those in \eqref{eq:bic_sim} for the FDM-LME, except that $b_{i,c-1}$ is replaced by $d_{i,c-1}$.

  \subsection{Simulation results}

    Due to space concerns, we have moved all the estimation results on the training data to Supplemental Materials, since prediction accuracy on testing data is generally considered more important than estimation accuracy on training data in degradation analysis. Our results show that both GPM and FDMs can estimate degradation amounts reasonably well. For the estimation of EODs, both FDM-LME and FDM-FLMM perform similarly with the one matching the true model doing slightly better than the other.
 
\begin{figure}[]
		\centering
		\includegraphics[width=0.7\linewidth]{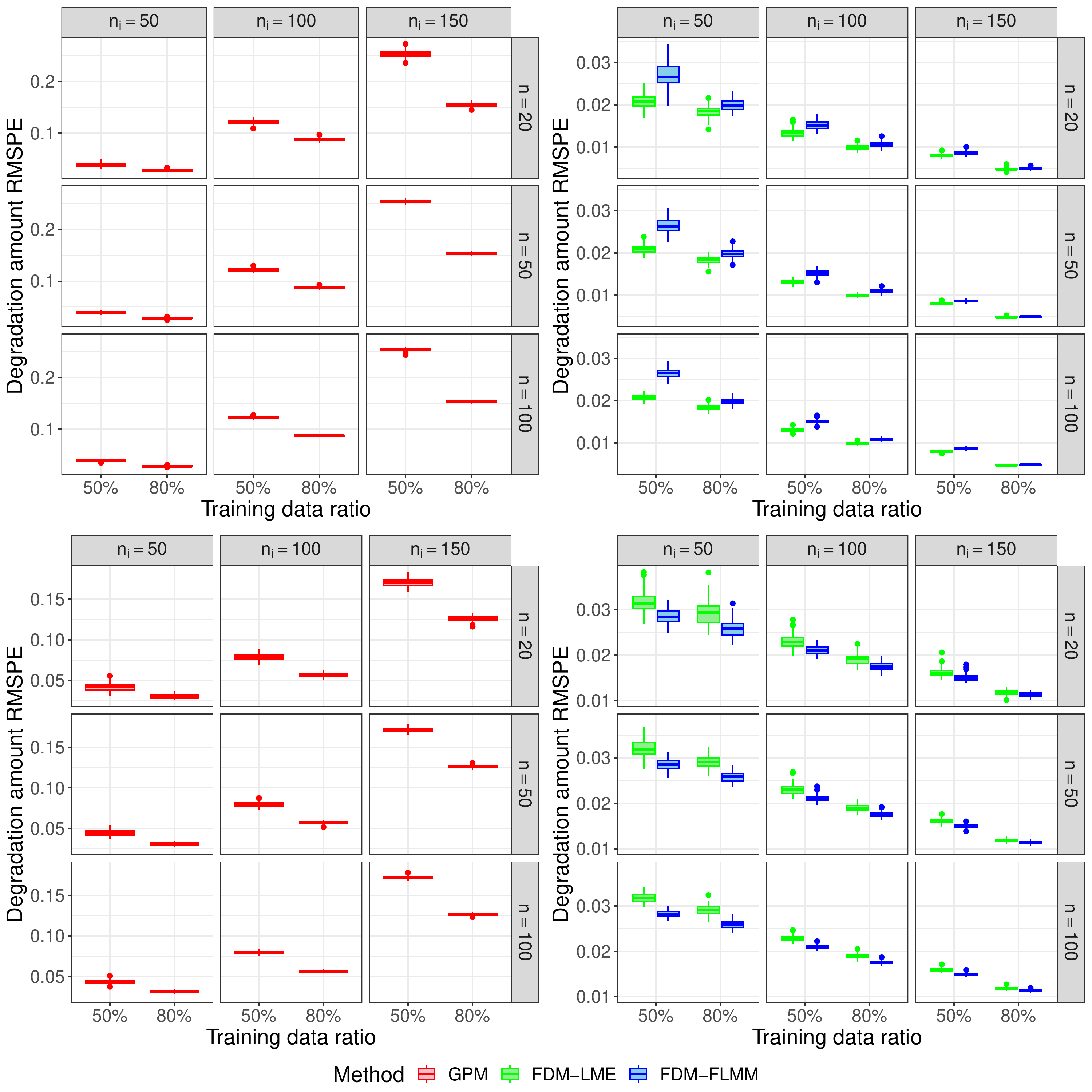}
		\caption{Degradation amount prediction performance comparison between GPM (left) and FDM (right). Top panels: simulation model \eqref{eq:bic_sim}; Bottom panels: simulation model \eqref{eq:bic_sim_flmm}. Note that left plots have much larger scales than right plots.}
		\label{fig:tst_deg}
	\end{figure}

	For prediction of degradation amounts, Figure \ref{fig:tst_deg} shows the boxplots of $\textrm{RMSPE}$s of the three methods grouped by simulation settings. It is evident that regardless of the scenario, FDM-LME consistently exhibits the lowest degradation RMSPE when EODs are generated by \eqref{eq:bic_sim}, while FDM-FLMM performs the best when EODs are generated by \eqref{eq:bic_sim_flmm}. Both FDM-LME and FDM-FLMM consistently outperform GPM by significant margins. Interestingly, when the number of cycles $n_i$ increases GPM appeared to overfit even more at the training step and thus suffered more at prediction. Note that increasing $tr\%$ generally leads to improved degradation RMSPEs for all methods across all scenarios. As $n_i$ increases, FDM methods performs much better than GPM, but FDM-LME and FDM-FLMM tend to have similar performance regardless of whether EODs are generated by \eqref{eq:bic_sim} or \eqref{eq:bic_sim_flmm}.

\begin{figure}[]
		\centering
		\includegraphics[width=0.7\linewidth]{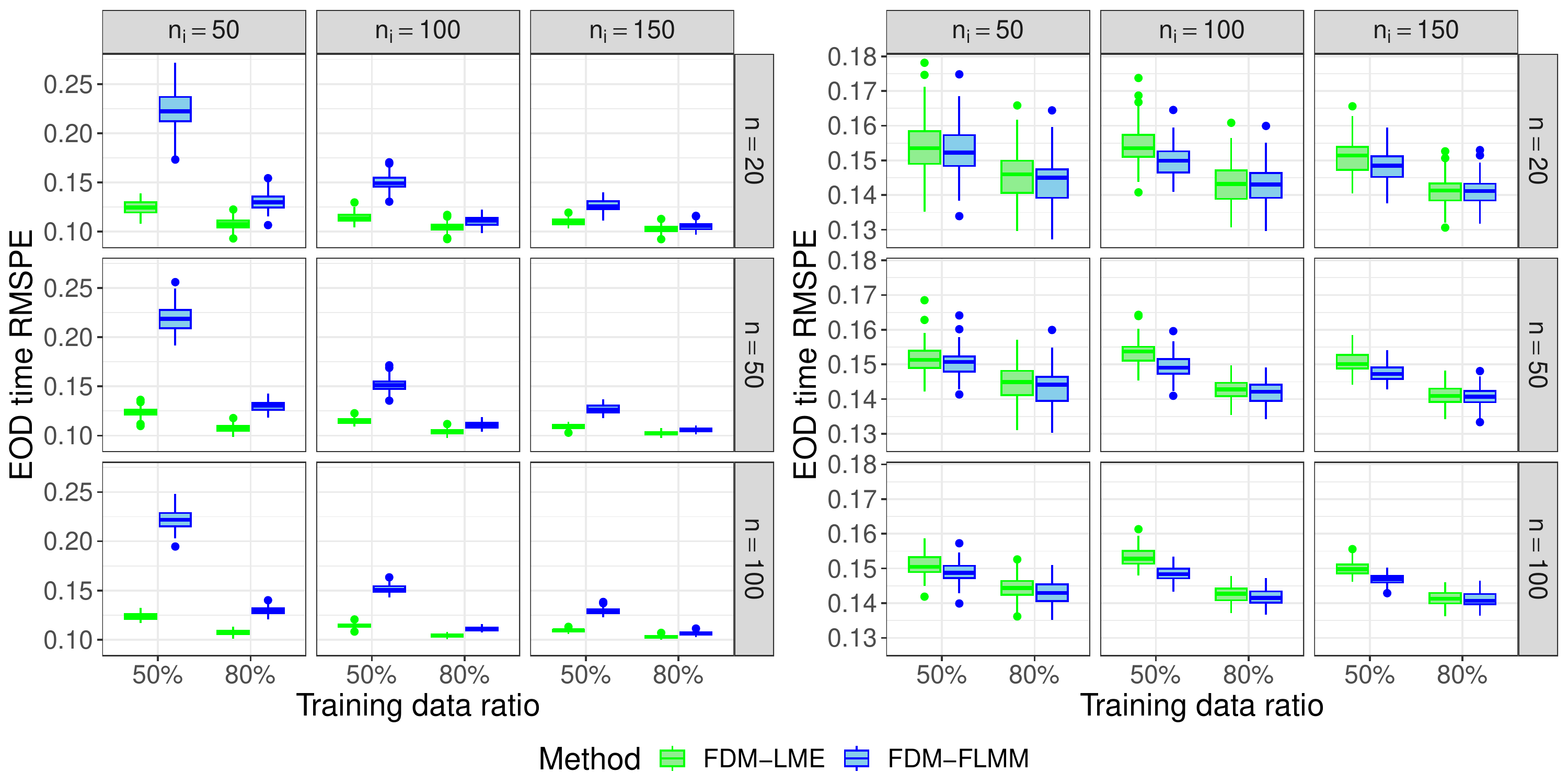}
		\caption{ EOD prediction performance comparison between FDM-LME and FDM-FLMM. Left: simulation model \eqref{eq:bic_sim}; Right: simulation model \eqref{eq:bic_sim_flmm}.}
		\label{fig:tst_eod}
	\end{figure}

Figure \ref{fig:tst_eod} compares the EOD prediction performances between FDM-LME and FDM-FLMM. Clearly, the method with a matching model generally predicts better than the other mis-matched one. But the difference becomes negligible when more training data, with increased $n_i$ and training data ratio, are used. 

\begin{figure}[]
		\centering
		\includegraphics[width=0.35\linewidth]{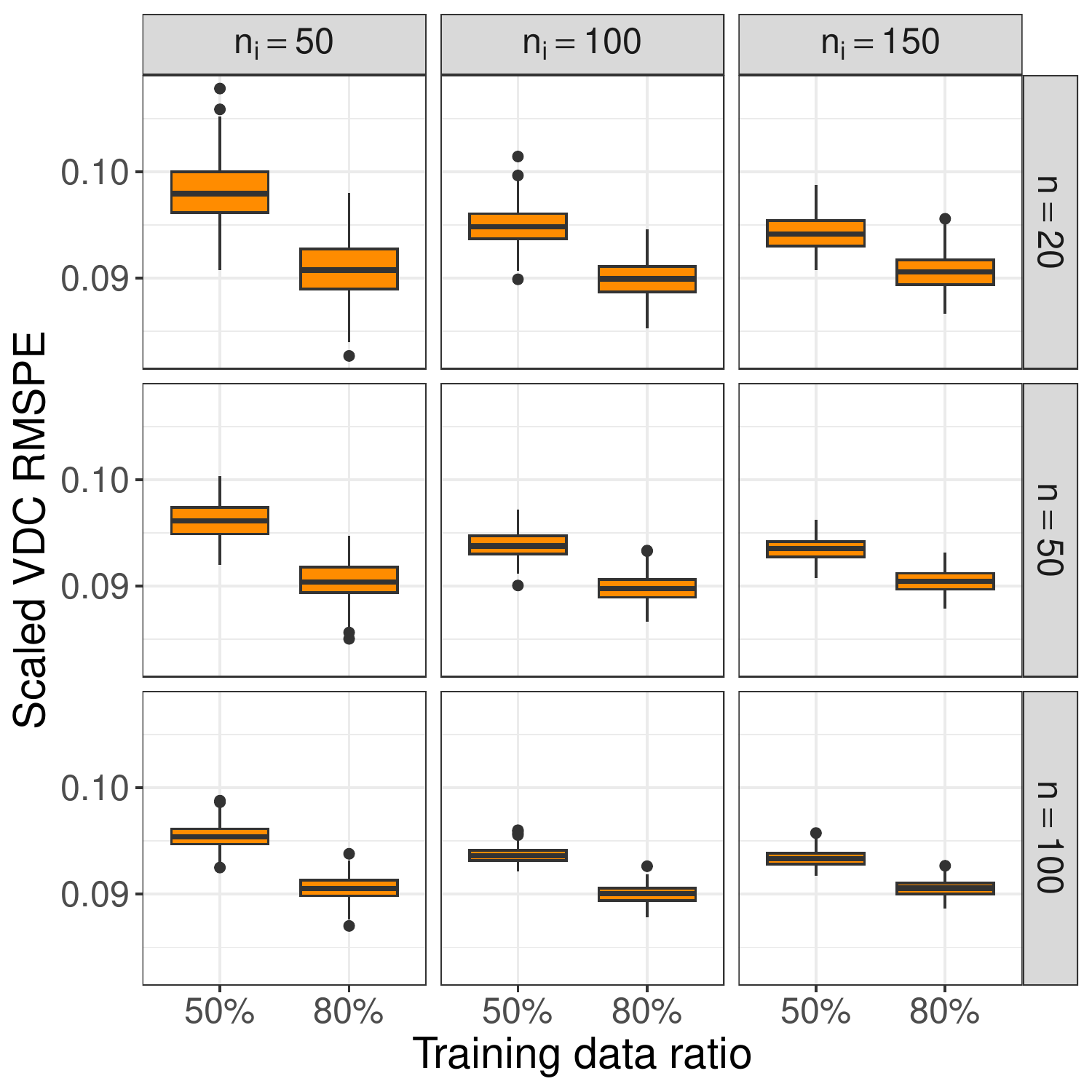}
		\caption{Scaled VDC prediction for FDM models in simulations based on model \eqref{eq:bic_sim_flmm}}
		\label{fig:tst_std_curve}
	\end{figure}

The prediction result of scaled VDCs by FDM models is plotted in Figure \ref{fig:tst_std_curve}. The prediction performance clearly improves when $n_i$ or the training data ratio increases.

	\section{Data Analysis}\label{se:analysis}
	
	\subsection{Model construction}\label{ss:da_model}
	
	The dataset contains voltage discharge curves for $n=20$ batteries under various experimental conditions. The experimental conditions are respectively the testing temperature (\emph{temp}) in Celsius degrees, discharge current (\emph{dc}) in Amp, and the stopping voltage (\emph{sv}) in volts. There are missing cycles for most batteries, in which case we re-indexed the $n_i$ available cycles for battery $i$ such that their cycle numbers would become 1 to $n_i$. The number of cycles $n_i$ for a battery ranges from 60 to 200. For each discharge cycle, the end of discharge (EOD) $b_{ic}$ is defined to be the time elapsed from the beginning to the end of the cycle. To obtain the scaled curves $x_{ic}(t)$, we scale the timestamps on each voltage discharge curve (VDC) $y_{ic}(r)$ by the corresponding $b_{ic}$. Since the time grid for the scaled curve $x_{ic}(t)$ may differ between cycles and batteries, we interpolate each curve to obtain observations on a uniform grid of 300 equally-spaced points. All the covariates are modeled as continuous variables. We apply the Arrhenius transformation for temperature using room temperature as the baseline level \citep{meeker2004degradation}, that is,
	$$
	z_1 = \frac{11604.52}{\text{\emph{temp}}^\circ \text{C} + 273.15} - \frac{11604.52}{24^\circ \text{C} + 273.15},
	$$
	where the denominators are essentially translations of temperatures in Celsius to Kelvin. 
	For discharge current (\emph{dc}) and stopping voltage \emph{sv}, we use the following transformations,
	$z_2 = {\log(\text{\emph{dc}})}/{2}$ and $z_3 = \log(\emph{sv})/{2}$.
	This is called the power law transformation using 2 Amps as baseline level for discharge current and 2 Volts as baseline level for stopping voltage \citep{meeker2004degradation}. These are standard transformations in battery studies and can enhance the numerical stability in model fitting. Furthermore, the baseline levels for the three experimental conditions have a covariate value of 0 under these transformations. This makes it easier to interpret the intercept of the model and compare the effects on the response between different levels of each covariate.
	
	When performing the FPCA on the scaled VDCs, we find that the first three FPCs can already explain more than $99\%$ of the total variation in the curves. Therefore, we use $K=3$ here and represent each curve $x_{ic}(t)$ by the three corresponding FPC scores, $\gamma_{ic,j}$, $j=1,2,3$. We fit the multivariate linear mixed-effects model in \eqref{eq:gamma} with $\zvec_{ic} = (z_{1i}, z_{2i}, z_{3i})^\top$, using the procedure introduced in Section \ref{se:model}.
	
	Once the model is fitted and the parameters are estimated, we apply
	\begin{equation}\label{eq:gammahat}
		\hat{\gammavec}_{ic^*} = \hat{\vvec}_0 + \hat{\uvec}_{0i} + (\hat{\vvec}_1 + \hat{\uvec}_{1i})c^* + \hat{\mathbf{P}}{\zvec_{ic^*}}
	\end{equation}
	to obtain the predicted FPC scores of the scaled voltage discharge curves $x_{ic^*}(t)$ for a future cycle $c^*$. Combining these scores with the FPCs and the estimated mean discharge function $\hat{\mu}(t)$ yields the prediction of the scaled curve $x_{ic^*}(t)$ for cycle $c^*$.
	
	The next step is to model the EOD time $b_{ic}$. Visual inspection in Figure \ref{fig:degradation} indicates that a linear trend in cycle number can capture well the trajectory of EOD time. To capture the variations between batteries, a reasonable model choice is the linear mixed-effects model with random terms for the intercept and slope for cycle number grouped by battery. Two additional covariates are introduced following the suggestion in \citet{saha2009pf}. One is the EOD $b_{i,c-1}$ of the previous cycle, to capture the autoregressive nature of EODs found in their analysis. The other is a transformation of the rest period $\Delta_{ic}$ between two cycles, namely, $\exp(-1/\Delta_{ic})$. Hence the first option to model $b_{ic}$ is the form of model \eqref{eq:bic_lme} with $\zvec_{1,ic}=(c,b_{i,c-1})^\top$ and $\zvec_{2,ic} = (\exp(-1/\Delta_{ic}),z_{1i}, z_{2i}, z_{3i})^\top$. 
	
	A second option is to take advantage of the information from scaled voltage discharge curve by fitting the functional linear mixed-effects model \eqref{eq:bic_flmm}. Since the experimental conditions have already been incorporated into the model for $x_{ic}(t)$, we choose not to include them again in the models for the EOD time, and thus $\{\zvec_{1,ic},\zvec_{2,ic}\}$ in \eqref{eq:bic_flmm} becomes $\zvec_{1,ic}=c$ and $\zvec_{2,ic} = (b_{i,c-1}, \exp(-1/\Delta_{ic}))^\top$. 
 We first fit models \eqref{eq:bic_lme} to \eqref{eq:bic_flmm} using the training data, that is, the first $tr\%$ of data for each battery. Based on the fitted models, we make predictions on the $1-tr\%$ held-out testing data.
	
	After prediction of the scaled curve $x_{ic}(t)$ as well as domain end time $b_{ic}$, we reconstruct the curve on its natural domain $y_{ic}(s)$ and compute the degradation amount \eqref{eq:dic} using the $L^p$ norm with $p=1$. As done in simulations, we compare our degradation predictions to those from the general path model which is based only on the degradation amounts at the training cycles.
	
	\subsection{Model evaluation}\label{ss:da_eval}
	\begin{figure}[H]
		\centering
		\includegraphics[width=0.6\linewidth]{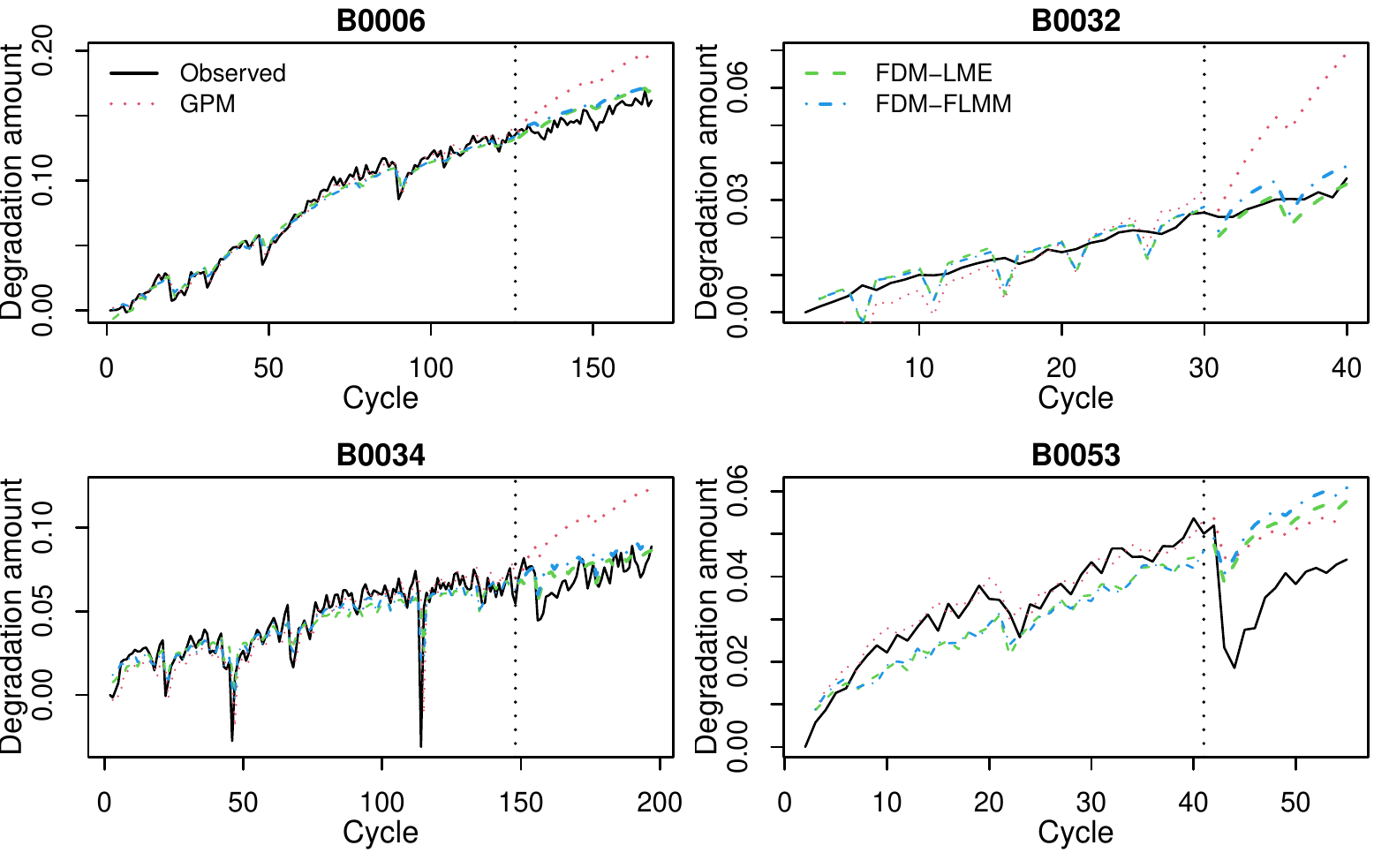}
		\caption{Degradation paths for 4 representing batteries overlayed with fitted and predicted paths by GPM, FDM-LME, and FDM-FLMM. Dotted vertical line indicate the separation between train (first $75\%$ of discharge cycles from each battery) and test set (last $25\%$ of discharge cycles from each battery).}
		\label{fig:degrad_preds75}
	\end{figure}	
 	\begin{figure}[H]
		\centering
		\includegraphics[width=0.6\linewidth]{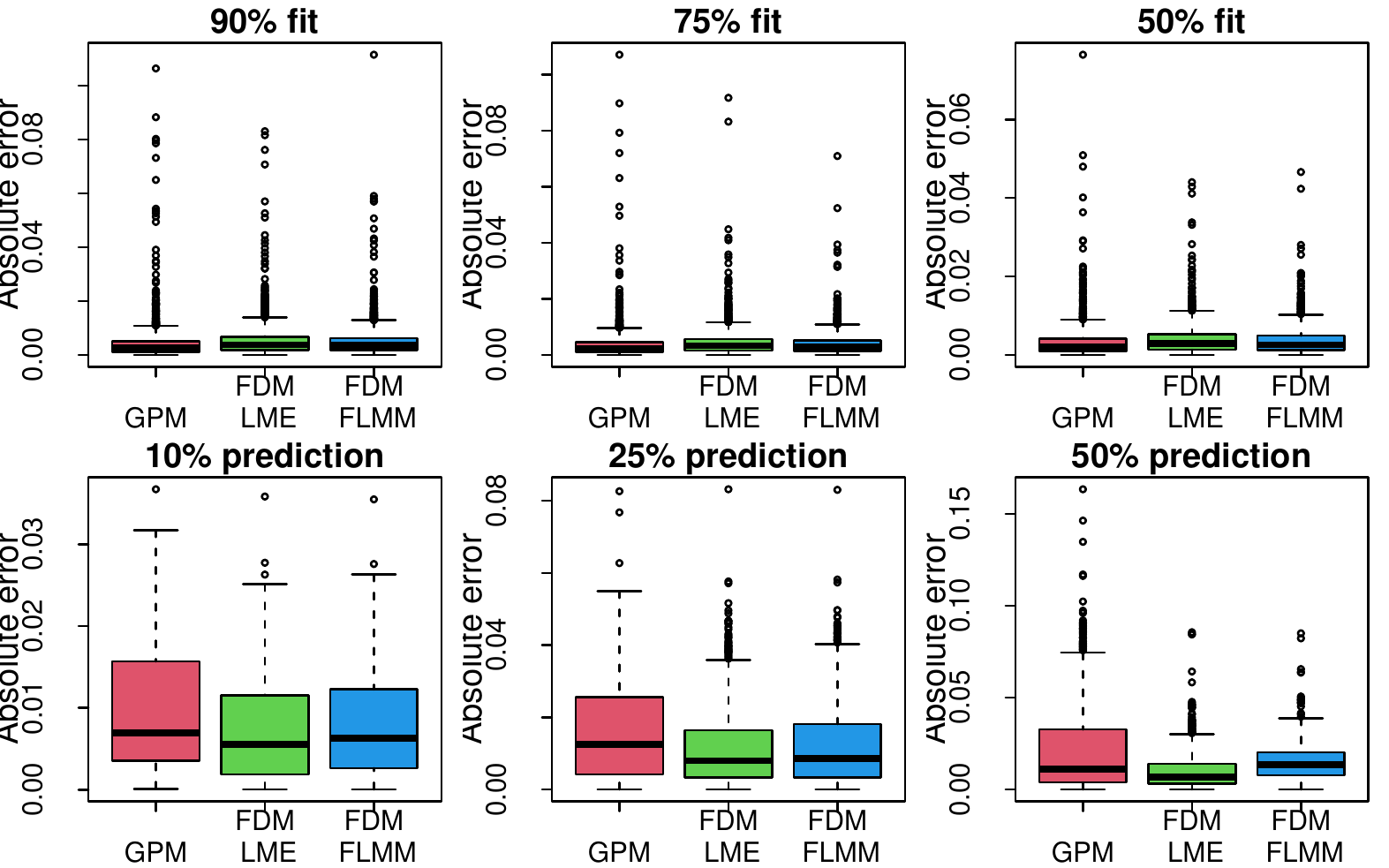}
		\caption{Boxplots of absolute error in degradation amount fit (top) and prediction (bottom) among three considered methods across three train/test split ratios.}
		\label{fig:degrad_rmses}
	\end{figure}

	We consider three levels $\{50\%,75\%,90\%\}$ for the train ratio $tr\%$. For illustration, we present the fitted and predicted degradation paths for the $75\%$ train ratio setting. Figure \ref{fig:degrad_preds75} displays the predicted degradation paths for 4 presenting batteries when $75\%$ of the available data for each battery are used in modeling. Predictions from FDM-LME and FDM-FLMM are able to capture well the testing part trajectories of the degradation paths for most batteries while GPM tends to overestimate degradation amounts. Prediction for battery B0053 is an exception with all models performing badly. This is because of the big dip in the degradation path corresponding to a longer than normal resting period $\Delta_{ic}$. While all three models have built-in mechanisms to capture this resting period, its effect is much more profound than the prediction from any model.

	Figure \ref{fig:degrad_rmses} shows boxplots of absolute errors of the three methods when fitting and predicting degradation amounts under the three train/test ratios. The fits on the training data from all three methods are comparable between each other. However, when comparing predictions, both FDMs deliver better and more consistent performance than the GPM. Since the FDM-LME exhibits the best degradation prediction performance in test dataset across all $tr\%$, we choose to use it as our final model for all the further data analysis.

	\subsection{Degradation prediction}\label{ss:da_fulldata}
	
\begin{figure}
		\centering
		\includegraphics[width=0.55\linewidth]{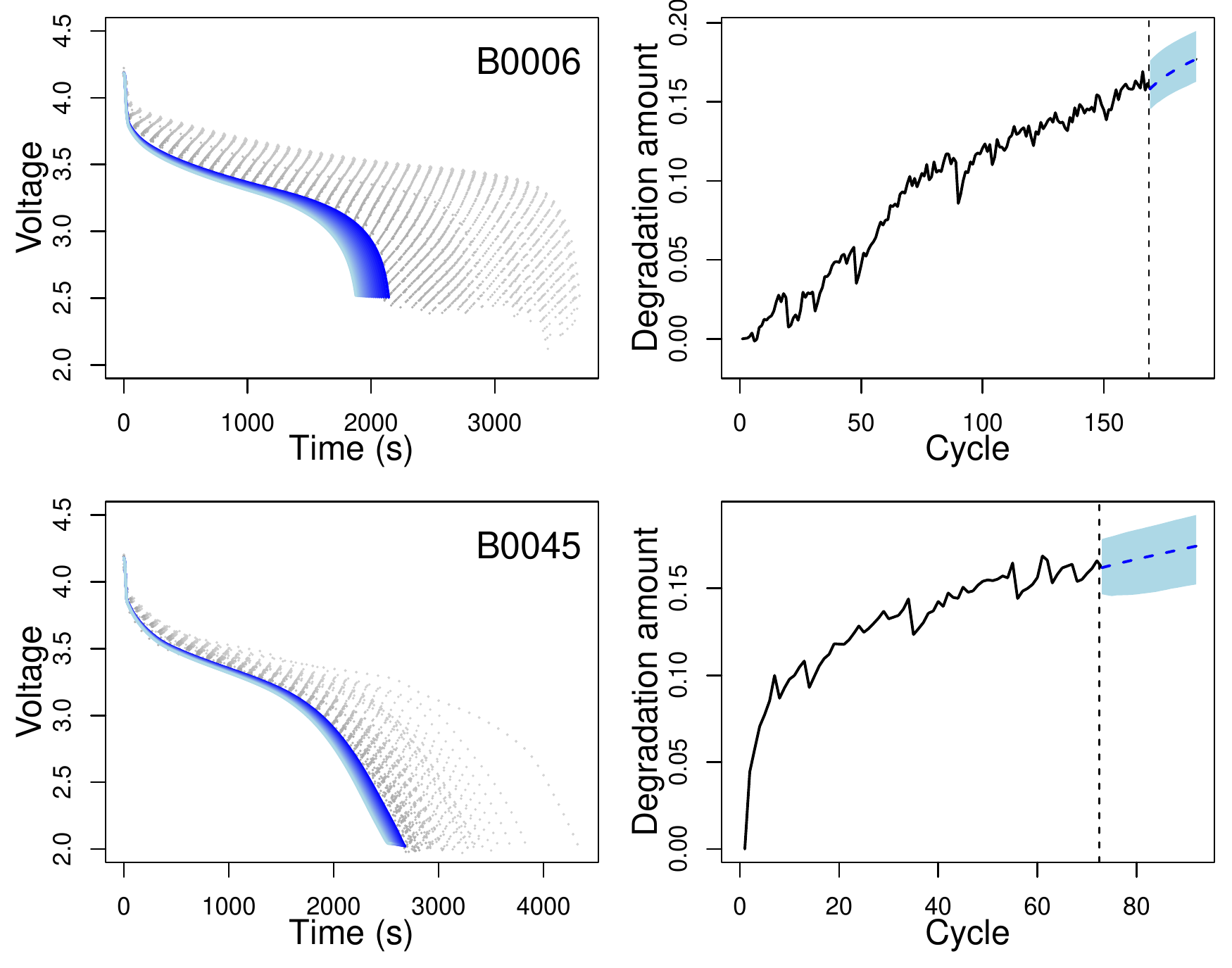}
  \includegraphics[width=0.16\linewidth]{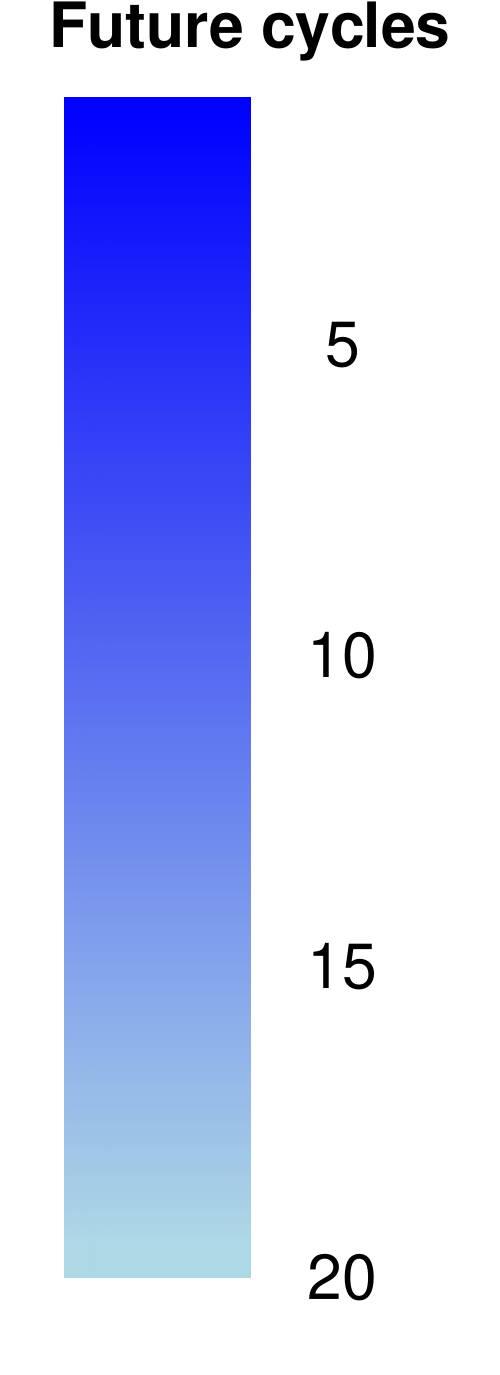}
		\caption{Predicted VDCs and degradation paths for two batteries in the next 20 discharge cycles based on FDM-LME. Left: Observed (dotted curves) and predicted VDCs (solid curves). Right: observed (solid lines) and predicted degradation paths (dashed lines) with $95\%$ degradation prediction intervals (shaded areas). Dashed vertical lines separate observed cycles from future cycles.}
		\label{fig:whole_data}
	\end{figure}

	We perform functional degradation analysis using all the available data on the batteries, that is, we make prediction about the battery degradation beyond the discharge cycles available in the data. Utilizing the outcomes derived from the FDM-LME model, which is our final model, we predict the VDCs and degradation paths for the next 20 cycles of each battery assuming a 5-hour lag between cycles. Additionally, we calculate the $95\%$ degradation prediction intervals (Section \ref{se:PI}). Figure \ref{fig:whole_data} visualizes the results for two batteries. This result illustrates the benefits of functional degradation model in producing both the predictions for VDCs and the degradation amount of interest. The estimation result for the training data is relegated again to the Supplemental Materials (Sections \ref{se:FPCA-full}, \ref{se:FDM-LME-full}, and \ref{se:FDM-FLMM-full}).
	
\subsection{Degradation prediction intervals}\label{se:PI}

As a two-step procedure, the FDM has multiple sources of uncertainties. At the first step of modeling the scaled VDCs, both the translation of the curves to their FPC scores and the multivariate mixed-effects model for the FPC scores can introduce estimation errors. At the second step, both LME and FLMM have their intrinsic uncertainty in modeling the EODs. The FLMM has an extra source of uncertainty from using the estimated $x_{ic}(\cdot)$ as a functional covariate. Given all these uncertainty resources, a rigorous theoretical assessment of the uncertainty is a daunting, if not impossible, task. Here, we consider the fractional random weight bootstrap method in \citet{hong2020boots} to construct prediction intervals for degradation amount. The detail of this procedure is in Section \ref{se:PIWboot} of the Supplementary Materials. Figure \ref{fig:degrad_PI75} presents the predicted degradation paths and $95\%$ degradation prediction intervals of the FDM-LME model for four selected batteries when $75\%$ of the available data for each battery are utilized in modeling. The proposed prediction interval covers degradation paths well for future cycles.

 	\begin{figure}[H]
		\centering
		\includegraphics[width=0.55\linewidth]{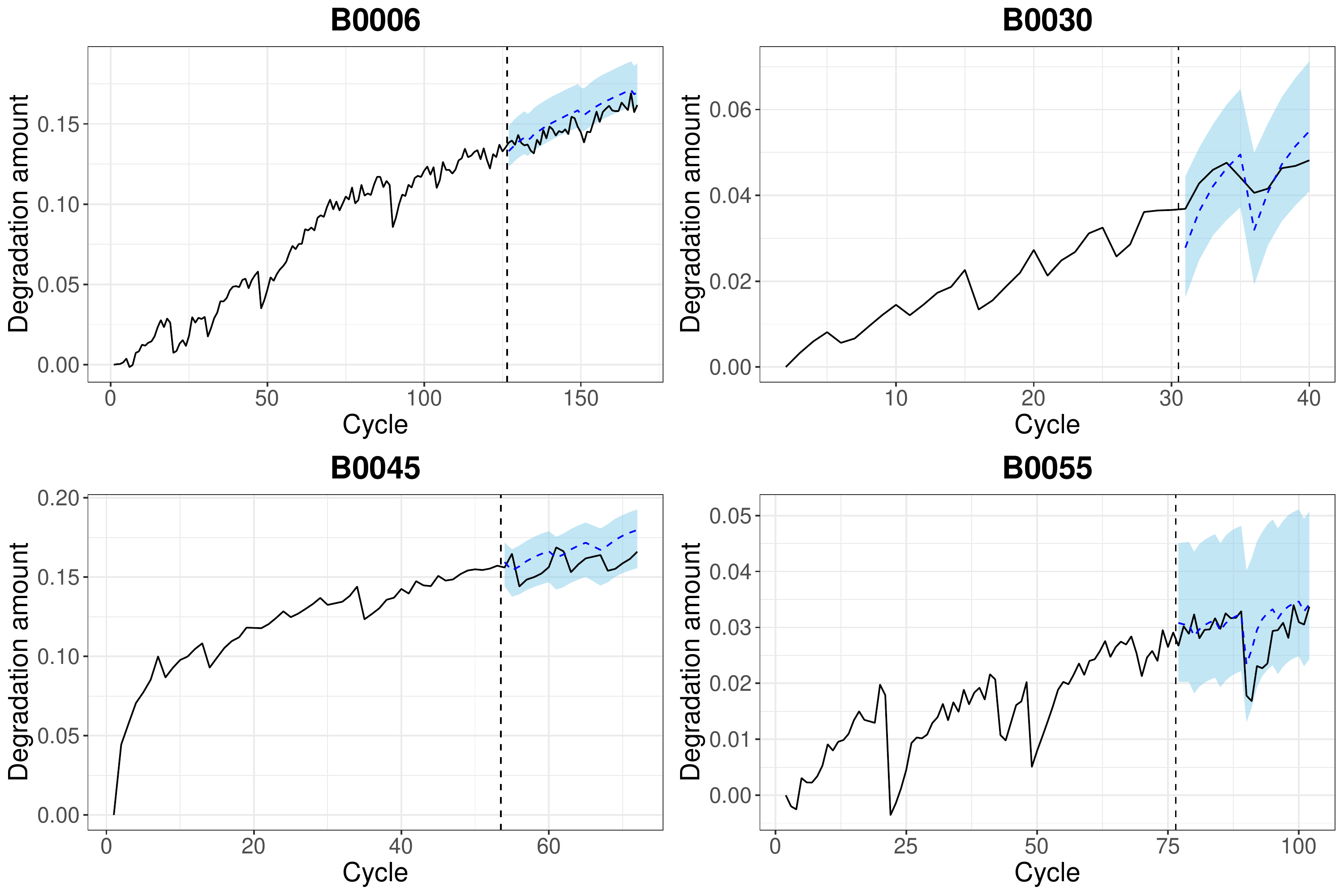}
		\caption{Degradation predictions and $95\%$ prediction intervals for 4 selected batteries. Plotted for each battery are the observed degradation path (solid line), the predicted degradation path (dashed line), and the degradation prediction intervals (shaded area). The dashed vertical line separates the train (first $75\%$ of discharge cycles) and test sets (last $25\%$ of discharge cycles).}
		\label{fig:degrad_PI75}
	\end{figure}

	\section{Conclusion and Future Work}\label{se:conclusion}
	
	In this work, we propose a functional degradation model to analyze a battery life testing study, where voltage discharge curves form a longitudinal set of functional data observations showing a pattern of degradation. Instead of applying standard degradation models directly to degradation measures summarizing functional observations, our functional data approach takes full advantage of the whole VDCs. Considering the heterogeneity in the supporting domains of the VDCs, we first re-scale each VDC to reside on the common domain $[0,1]$ and then simultaneously model the scaled curves and the EOD times through mixed-effects models. Simulations show that our functional degradation model can indeed yield better predictions than the standard general path model. Our detailed analysis of the battery life testing data further reveal the great potential of the functional degradation model for more accurate and more stable degradation prediction for Lithium-ion batteries. Another advantage of the functional approach is that its prediction is not only just a predicted value for the degradation amount but also a complete VDC. The availability of complete VDC predictions would allow the practitioners the flexibility of defining their own degradation measures based on the curves.
	
	There are, however, potential assumptions in our models that may not be realistic and should be studied in future work. First, we assume a linear effect from the experimental conditions. This might not be true from physical sciences especially with factors like temperature. The normality assumption for random errors and random effects can also be restrictive. Relaxation of such assumptions are interesting topics to explore in the future.
	
	The degradation paths show an interesting impact from resting periods and accounting for the resting time helps improve local and overall prediction accuracy. Currently, in order to predict degradation in the future, resting time needs to be preset. This is not realistic. A better approach would be to account for randomness in the resting amount. As seen in Figure \ref{fig:degrad_preds75}, a long resting period in battery B0053 completely drives the data away from all models' prediction. It would be beneficial to develop a guide on when to reset model training based on length of resting period.  In addition, it will be interesting to consider time-varying covariates such as varying temperature, multiple current loading values, and stopping voltage levels that interchange across discharge cycles. The latter would be for a more realistic situation where batteries are discharged to half or a quarter of its voltage capacity instead of a full discharge like our current application.

 Our proposed functional degradation modeling framework can be extended to other applications with similar data structure. In models \eqref{eq:bic_lme} and \eqref{eq:bic_flmm}, a linear trend is used to describe the trend over cycles $c$, for our battery application. Such simple linear trend works well in our battery application. For the other data with flatter-towards-the-end behavior, some modifications of the model are necessary. In the degradation literature, there is a wide class of functional forms that can be used to fit those kinds of behavior (e.g., the S shaped functional form used in \citet{duan2017}). Of course, introducing the S shaped functional form of cycle effect will complicate the estimation procedure. This is a very interesting point for future research.

\begin{acks}[Acknowledgments]
Youngjin Cho and Quyen Do are joint first authors.
The authors are grateful to the Editor, the Associate Editor and three  anonymous referees for their constructive comments that have significantly improved the quality of this paper.
The authors acknowledge the Advanced Research Computing program at Virginia Tech for providing computational resources and the NASA Ames Prognostics Data Repository for providing the battery dataset. 
\end{acks}

\begin{funding}
Du's research was partly supported by the U.S. National Science Foundation Grant DMS-1916174. The work by Hong was partially supported by the U.S. National Science Foundation Grant CMMI-1904165 to Virginia Tech.
\end{funding}

\begin{supplement}
\stitle{Supplement to ``Reliability Study of Battery Lives: A Functional Degradation Analysis Approach''.}
\sdescription{The supplement contains the additional results in simulation study and data analysis.}
\end{supplement}


\bibliographystyle{imsart-nameyear} 
	\bibliography{aoasref}

\newpage
\newpage
\setcounter{page}{1}
\begin{title}
	\centering
	\Large Supplement to ``Reliability Study of Battery Lives: A Functional Degradation Analysis Approach'' \par
	\vskip 1em
	\large {\centering
		Youngjin Cho, Quyen Do, Pang Du, and Yili Hong\\[1.5ex]
	} \par
	\large {\centering
		Department of Statistics, Virginia Tech, Blacksburg, VA 24061;\\[1.5ex]
	} \par
	\large {\centering
		Corning Inc.\\[1.5ex]
	} \par
\end{title}

\begin{appendix}
\counterwithin{figure}{section}

\section{Additional results in simulation study}\label{se:simul_supp}

Here we present additional results from the simulation study in Section \ref{se:simulation}. The RMSPE and RMSE for scaled VDCs are defined as $$\sqrt{\frac{\sum_{i=1}^n\sum_{c=\tilde{n}_i}^{n_i}\int_0^1(\hat{x}_{ic}(t) - x_{ic}(t))^2dt}{\sum_{i=1}^n(n_i-\tilde{n}_i+1)}} \mbox{ and } \sqrt{\frac{\sum_{i=1}^n\sum_{c=1}^{\tilde{n}_i - 1}\int_0^1(\hat{x}_{ic}(t) - x_{ic}(t))^2dt}{\sum_{i=1}^n(\tilde{n}_i-1)}}.$$  The RMSPE and RMSE for EODs are defined as $$\sqrt{\frac{\sum_{i=1}^n\sum_{c=\tilde{n}_i}^{n_i}(\hat{b}_{ic} - b_{ic})^2}{\sum_{i=1}^n(n_i-\tilde{n}_i+1)}} \mbox{ and }\sqrt{\frac{\sum_{i=1}^n\sum_{c=1}^{\tilde{n}_i - 1}(\hat{b}_{ic} - b_{ic})^2}{\sum_{i=1}^n(\tilde{n}_i-1)}}.$$ 

\begin{figure}[H]
		\centering
		\includegraphics[width=0.7\linewidth]{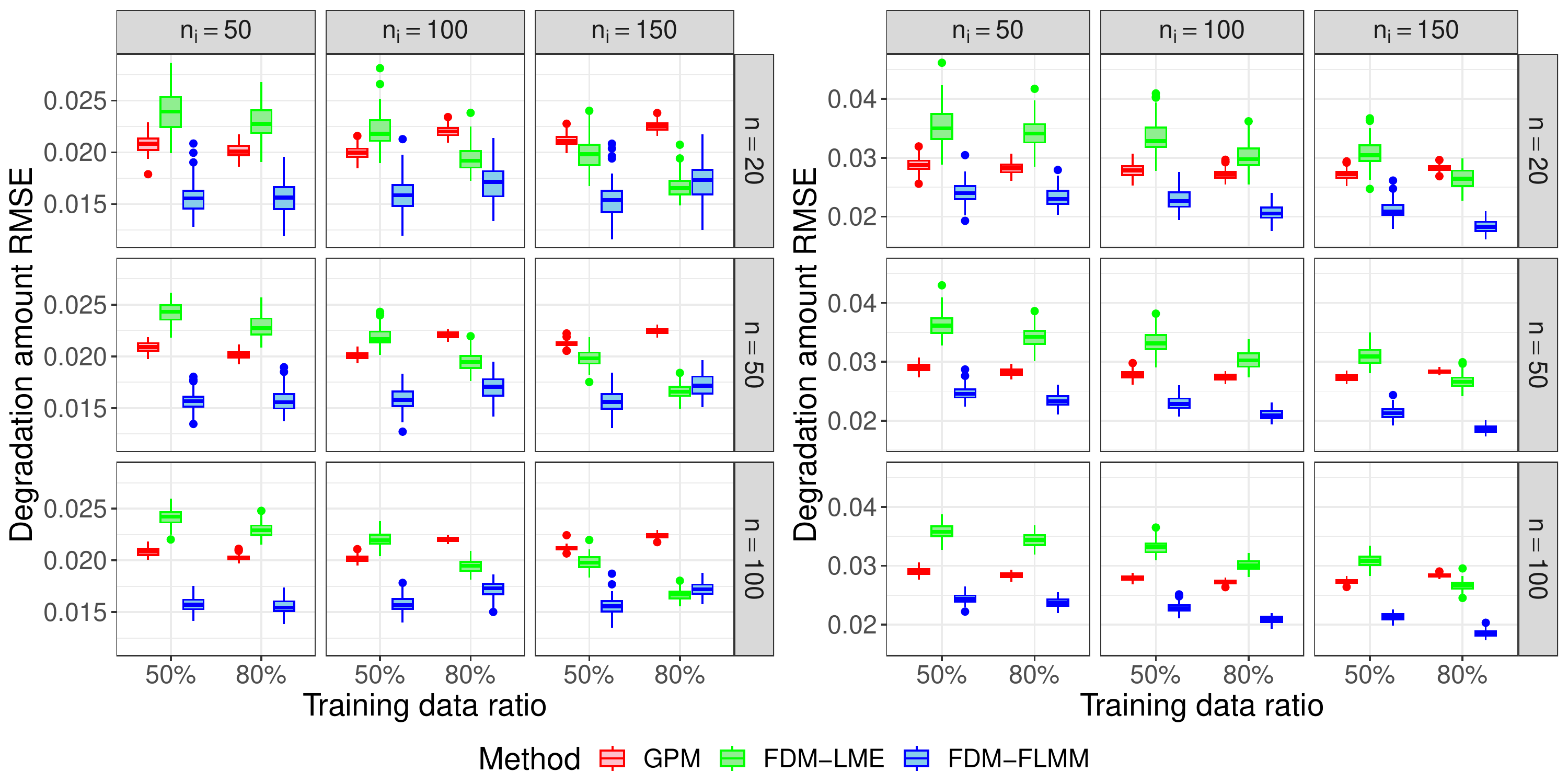}
		\caption{Degradation estimation performance on training data over all simulation settings. Left: simulation with model \eqref{eq:bic_sim}; Right: simulation with model \eqref{eq:bic_sim_flmm}.}
		\label{fig:tr_deg}
	\end{figure}

\begin{figure}[H]
		\centering
		\includegraphics[width=0.7\linewidth]{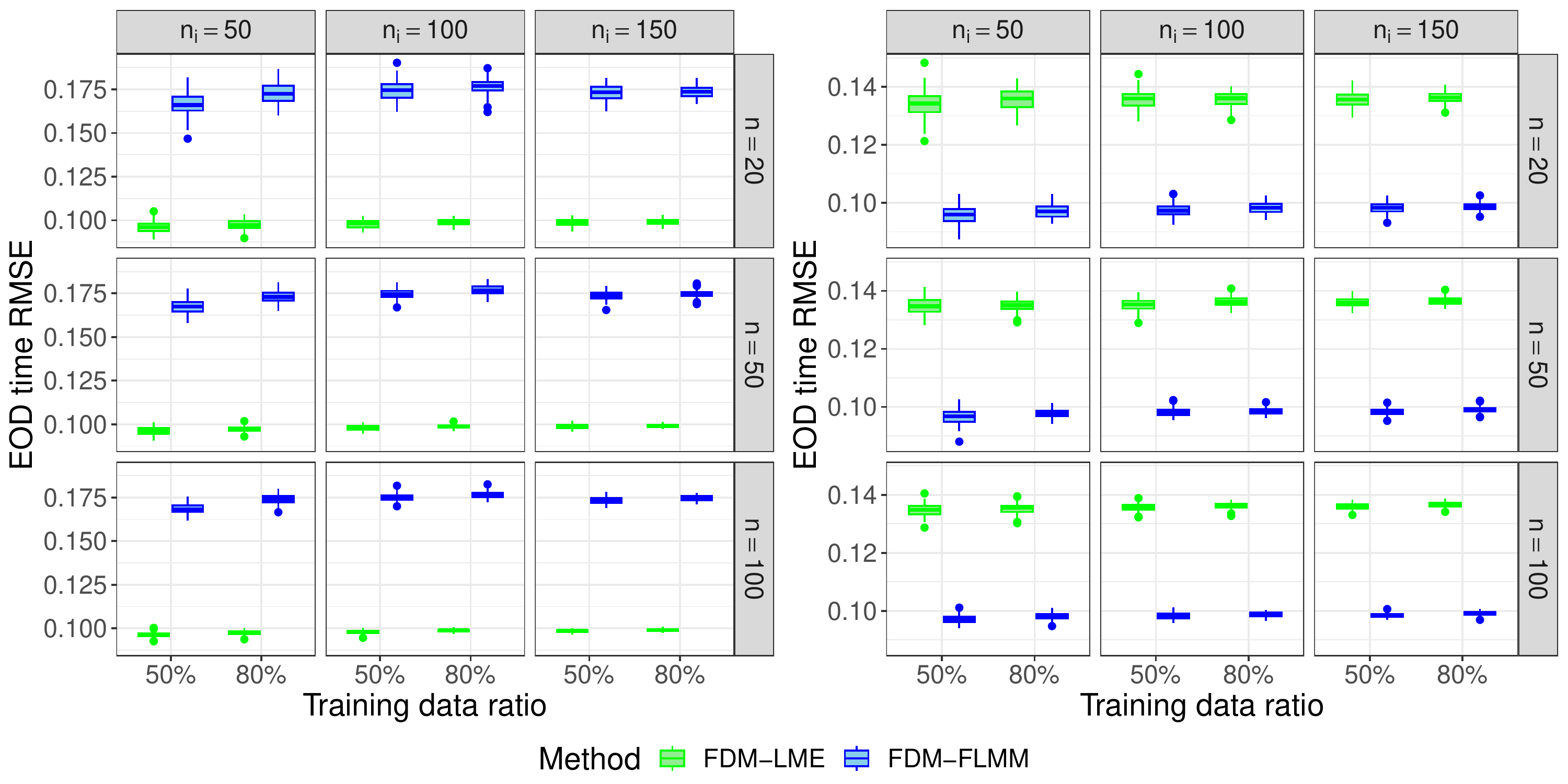}
		\caption{EOD fits performance on training data by FDM-LME and FDM-FLMM over all simulation settings. Left: simulation with model \eqref{eq:bic_sim}; Right: simulation with model \eqref{eq:bic_sim_flmm}.}
		\label{fig:tr_eod}
	\end{figure}

\begin{figure}[H]
		\centering
		\includegraphics[width=0.35\linewidth]{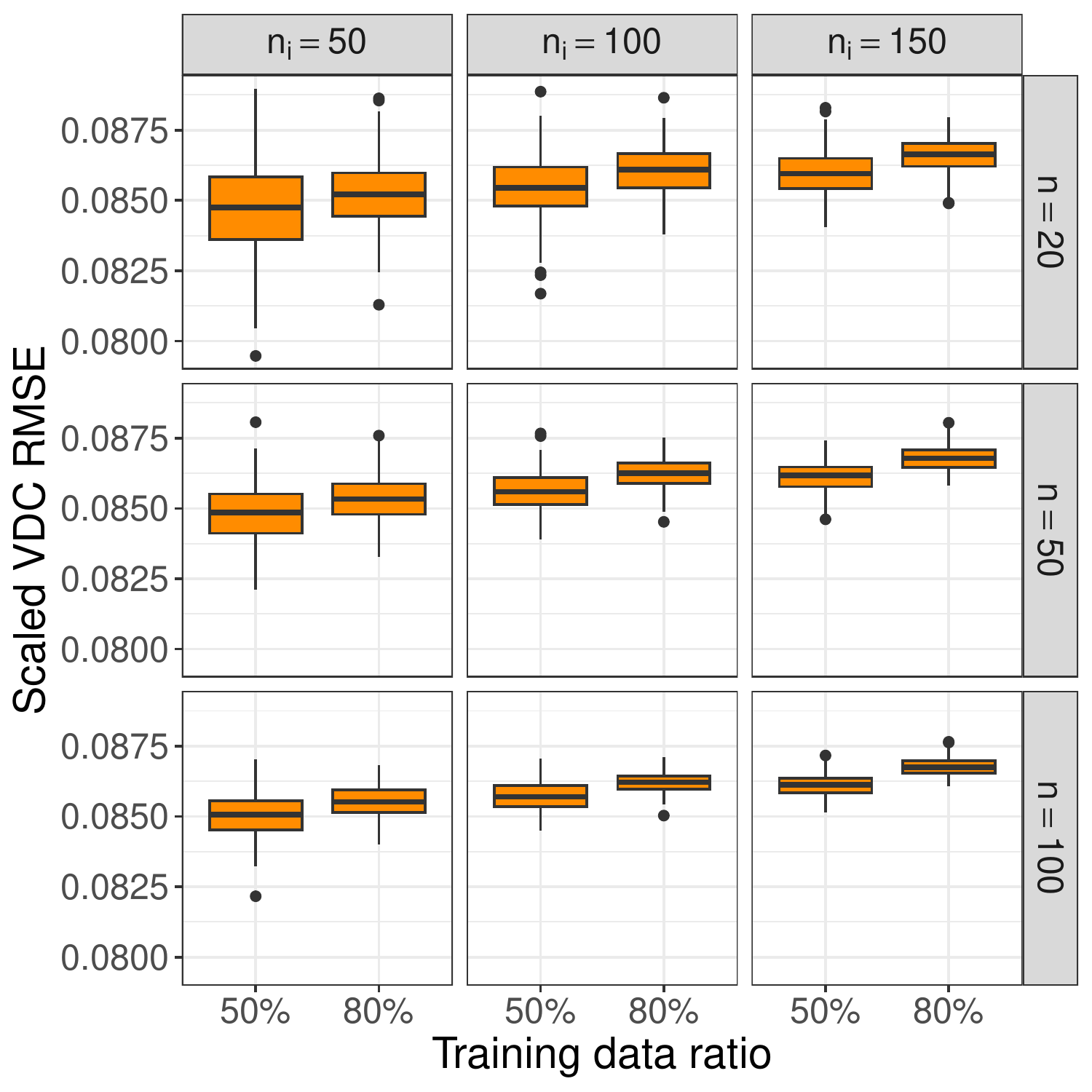}
		\caption{Fits of scaled VDCs by FDM models on the training data over all simulation settings.}
		\label{fig:tr_std_curbe}
	\end{figure}

\begin{figure}[H]
		\centering
		\includegraphics[width=0.7\linewidth]{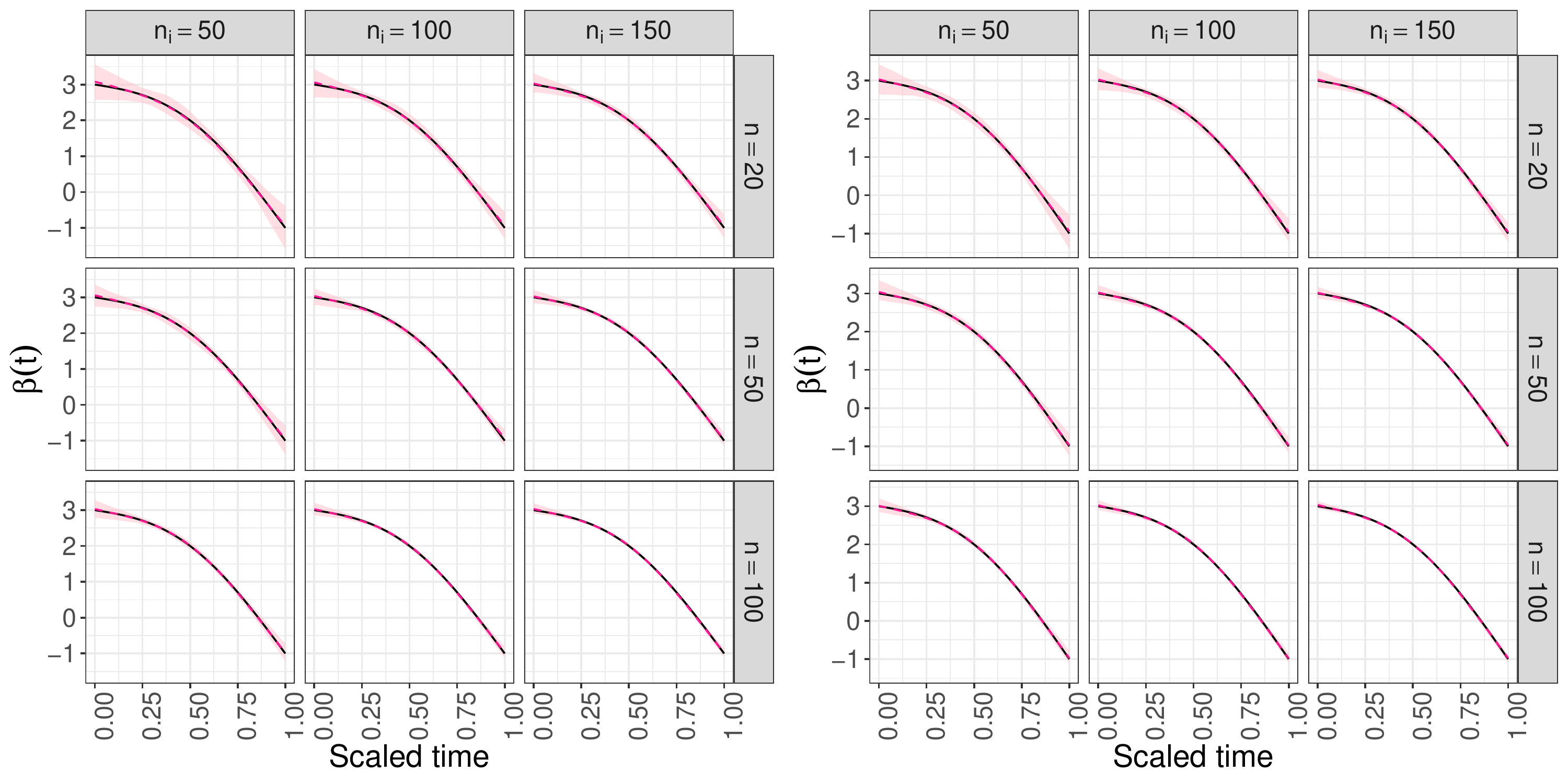}
		\caption{The performance of FDM-FLMM in estimating $\beta(\cdot)$ with simulation model \eqref{eq:bic_sim_flmm}. For each plot, the solid line is the true function $\beta(\cdot)$, the dahsed line is the pointwise average of $\hat{\beta}(\cdot)$ over 100 replicates, and shaded areas are pointwise $2.5\%$ and $97.5\%$ empirical percentiles of $\hat{\beta}(\cdot)$ over 100 replicates, respectively. $tr\%$ in left plots are $50\%$ and $tr\%$ in right plots are $80\%$.}
		\label{fig:slope_fit}
	\end{figure}

In Figures \ref{fig:tr_deg}, \ref{fig:tr_eod}, and \ref{fig:tr_std_curbe}, we present the estimation performances on training data in terms of $\textrm{RMSE}$s for degradation, scaled VDCs, and EODs. Figure \ref{fig:slope_fit} displays the performance of estimating $\beta(\cdot)$ using \eqref{eq:bic_flmm} when EOD in simulated data is generated according to \eqref{eq:bic_sim_flmm}.

Overall, for degradation, in Figure \ref{fig:tr_deg}, FDM-FLMM generally exhibits the smallest degradation RMSE, regardless of whether EOD is generated by \eqref{eq:bic_sim} or \eqref{eq:bic_sim_flmm}. FDM-LME tends to have larger RMSE compared to the other two methods, even when EOD is generated by \eqref{eq:bic_sim}. We observe that an increase in either $n_i$ or $tr\%$ leads to smaller RMSE for both FDM-LME and FDM-FLMM, resulting in FDM-LME occasionally having the smallest RMSE for certain scenarios. It is not surprising that GPM has reasonable performance in fitting the training data compared to the two FDM models. The GPM directly models the degradation amounts. Although this yields relatively good fit on training data, it also tends to overfit the data and thus leads to worse prediction performance as demonstrated above. On the other hand, the two FDM models deal with functional VDCs from two sources: the scaled VDCs and the EODs. Although the FDM approaches do not fit the training data as the direct GPM approach, resulting in not overwhelming training performances over GPM, they gain in the prediction performance which is more important in degradation analysis.

Figure \ref{fig:tr_eod} compares the fits of EODs on training data for the two FDMs. It is evident that the FDM matching the true model always fits better than the other FDM. Figure \ref{fig:tr_std_curbe} examines the fits of the scaled VDCs by FDM models. One can see that RMSEs for scaled VDCs are generally satisfactory.

The performances of FDM-FLMM in estimating $\beta(\cdot)$ when the EOD is generated by \eqref{eq:bic_sim_flmm} are presented in Figure 
\ref{fig:slope_fit}. One can see that FDM-FLMM accurately estimates the true slope function.

\section{Fractional random weight bootstrap for degradation prediction interval}\label{se:PIWboot}
Here we describe the method for computing prediction intervals for FDM-LME, the final model in the application, with the notion that such a method for FDM-FLMM will be similar. The method combines the nonparametric bootstrap prediction interval procedure with the fractional random weight bootstrap idea in \citet{hong2020boots}. 
Let the density function of the Dirichlet distribution of order $n$ with parameters $\{\lambda_i\}_{i=1}^n$ be
\begin{equation}\label{eqn:dirichlet}
    f(w_1,\cdots,w_n;\lambda_1,\cdots,\lambda_n)=\frac{1}{\Omega(\lambda_1,\cdots,\lambda_n)}\prod_{i=1}^n w_i^{\lambda_i-1},
\end{equation} where $\sum_{i=1}^n w_i=1$, $w_i \geq 0$ for $i=1,\cdots,n$, and $\Omega(\lambda_1,\cdots,\lambda_n)$ is the normalizing factor. Consider $B=5000$ bootstrap samples. Recall that we have $n$ batteries with battery $i$ having data on a total of $n_i-1$ cycles. For each battery $i$, we want to use the available $\tilde{n}_i-1$ cycles to make predictions about future cycles $\tilde{n}_i, \tilde{n}_i+1,\ldots,n_i$. The 95\% prediction intervals for these predictions are then computed as follows:

\begin{enumerate}
    \item For each $b=1, \ldots, B$, we iterate the following steps.
    \begin{enumerate}
        \item Generate $\tilde{\wvec}^{(b)}=(\tilde{w}^{(b)}_1,\cdots,\tilde{w}^{(b)}_n)^\top$ using the Dirichlet distribution \eqref{eqn:dirichlet} with $\lambda_i=1$ for all $i$ (i.e., all batteries are equally preferred). Scale $\tilde{\wvec}^{(b)}$ by $n$ to obtain $\wvec^{(b)}=n\tilde{\wvec}^{(b)}=(w^{(b)}_1,\cdots,w^{(b)}_n)^\top$ such that $\sum_{i=1}^nw^{(b)}_i=n$.
        \item Use weights $\wvec^{(b)}$ and $tr\%$ of the total cycles in each battery to fit {\it weighted linear mixed effects models (wLMEs)} \eqref{eq:gamma} and \eqref{eq:bic_lme} respectively for scaled VDCs and EODs. 
        \item For each $c=1, \ldots, n_i$, apply the fitted wLMEs to the covariates to obtain the fitted degradation amounts $\left\{\{\hat{d}_{ic}^{(b)}\}_{c=1}^{\tilde{n}_i-1}\right\}_{i=1}^n$ and the predicted degradation amounts $\left\{\{\hat{d}_{ic}^{(b)}\}_{c=\tilde{n}_i}^{n_i}\right\}_{i=1}^n$.
        \item Compute the residual degradation amounts $\left\{\left\{d_{ic}-\hat{d}_{ic}^{(b)}\right\}_{c=1}^{\tilde{n}_i-1}\right\}_{i=1}^n$ from the training portion of the cycles in the bootstrap sample, randomly sample an $e^{(b)}$ from them, and compute the degradation predictions for the $b$th bootstrap sample as $\left\{\{\hat{d}_{ic}^{(b)}+e^{(b)}\}_{c=\tilde{n}_i}^{n_i}\right\}_{i=1}^n$. 
    \end{enumerate}
    \item For each battery $i=1,\ldots,n$ and its future cycles $c=\tilde{n}_i, \tilde{n}_i+1, \ldots, n_i$, the 95\% prediction interval for the degradation amount $d_{ic}$ is then formed by the $2.5\%$ and $97.5\%$ empirical quantiles of $\left\{\hat{d}_{ic}^{(b)}+e^{(b)}\right\}_{b=1}^B$.
\end{enumerate}

\section{Functional principal components for scaled discharge voltage curves}\label{se:FPCA-full}
\begin{figure}[H]
		\centering
		\includegraphics[width=0.6\linewidth]{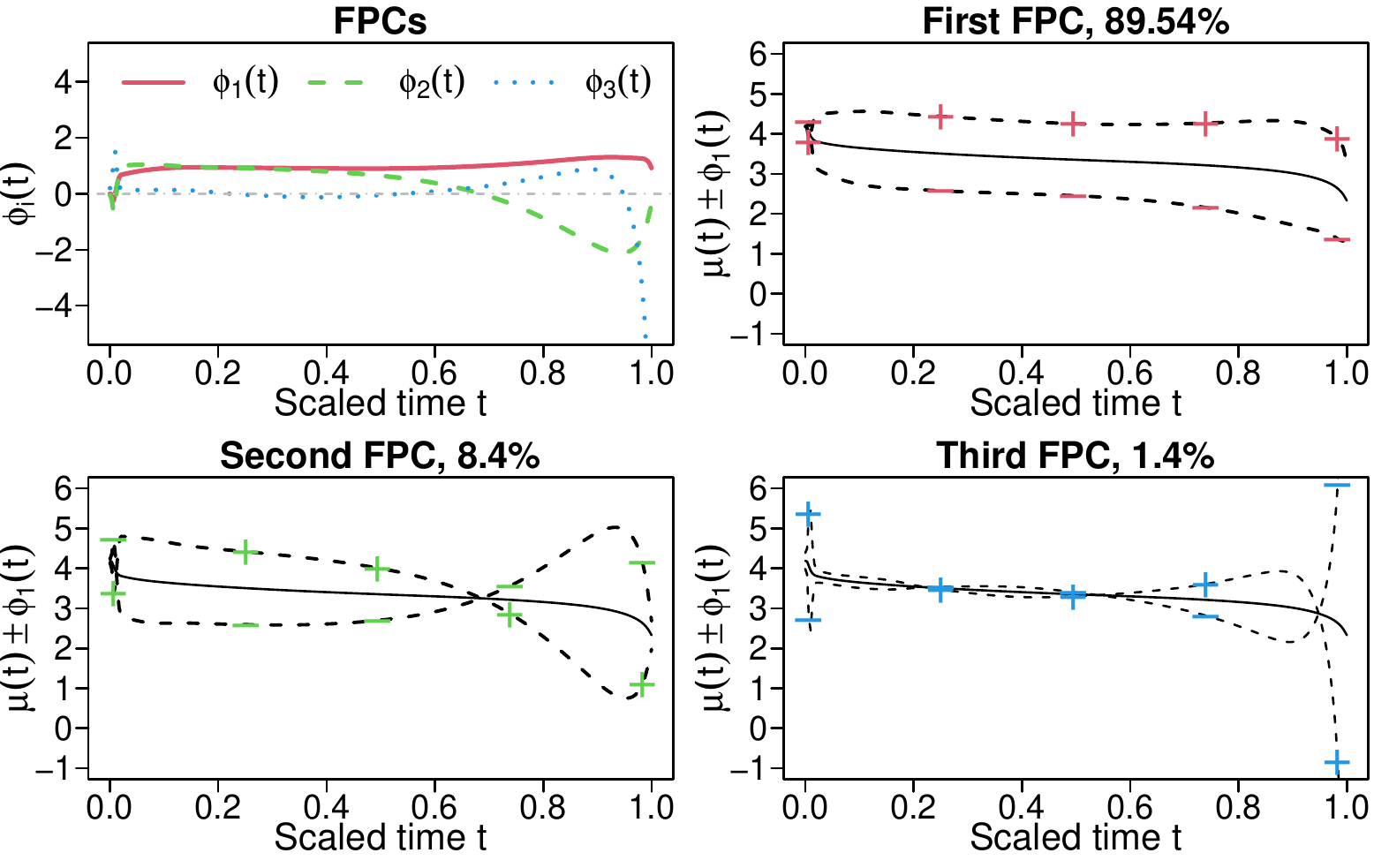}
		\caption{First three principal component functions with their proportion of total variation explained from FPC decomposition of scaled VDCs for available data from 20 batteries. Top left show the plots of the three principal components. Top right and the bottom two plots show the effect on the mean function (solid line) after adding and subtracting the FPCs.}
		\label{fig:fpcs}
	\end{figure}
	
	Figure \ref{fig:fpcs} visualizes the first three estimated functional principal components obtained from FPC decomposition of the scaled VDCs from all available discharge data. This decomposition result is used in Section \ref{ss:da_fulldata}. Visualizing the FPCs by adding and subtracting them from the mean curves helps illustrate their interpretation. The first FPC, accounting for $89.54\%$ of total variation in the data, is a positive function across the domain. Besides having values tapering toward 0 at the two ends, the first FPC is quite flat and its values consistent throughout the domain. The top right plot in Figure \ref{fig:fpcs} suggests that scaled VDC with positive score for the first FPC have higher voltage than the average scaled VDC almost entirely throughout the domain or discharge cycle. On the other hand, scaled VDC with negative score has lower voltage level than the average scaled VDC on the entire domain. Hence, first FPC indicates the deviation from the mean scaled VDC. The sign and magnitude of the first FPC score quantify the direction and strength of this deviation.
	
	The second FPC explains $8.4\%$ of total variation of the data. It has positive values approximately before $0.7$ mark and negative after that. It represents the difference in discharge voltage in the first $70\%$ of the cycle. This second FPC represents the contrast between voltage level in the first $70\%$ of the discharge cycle and the latter $30\%$. Scaled VDCs with high second FPC scores are those with higher than average voltage in the first $70\%$ of the discharge cycle but the voltage levels drop more significantly than the average in the remaining time of the cycle. Cycles with negative scores are those with lower voltage at the beginning of the cycle but the drop in later into the cycle is not as dramatic compared to the mean curve.
	
	The third FPC shows close to 0 values for the majority of the cycle except for the beginning and end of the cycle. This represents the variation in the scaled VDCs in terms of the beginning voltage level and how voltage drop at the end of the discharge cycles.

\section{Estimates from FDM-LME for EODs}\label{se:FDM-LME-full}
The FDM-LME for EOD $b_{ic}$ used in Section \ref{ss:da_fulldata} is  
\begin{multline}\label{eq:all_lme}
		b_{ic} = \alpha_0+w_{0i} + (\alpha_1+w_{1i})c + (\alpha_2+w_{2i})b_{i,c-1} + \alpha_3\exp(-1/\Delta_{ic}) \\+ \alpha_4 z_{1i} + \alpha_5 z_{2i} + \alpha_6 z_{3i} +\epsilon_{ic}.
	\end{multline}
 The estimates for the fixed-effects coefficients and their standard errors (se) are $\hat{\alpha}_0 = 382.254$ with se 136.816, $\hat{\alpha}_1 = -1.590$ with se 0.420, $\hat{\alpha}_2 = 0.695$ with se 0.036, $\hat{\alpha}_3 = 537.689$ with se 27.399, $\hat{\alpha}_4 = -58.898$ with se 23.385, $\hat{\alpha}_5 = -550.532$ with se 111.762, and $\hat{\alpha}_6 = 634.051$ with se 279.117. Note that all these fixed-effects coefficients are statistically significant under $95\%$ confidence level. This fitted FDM-LME was used in Section \ref{ss:da_fulldata} for the prediction of the future cycles for the batteries.

\section{Estimates from FDM-FLMM for EODs}\label{se:FDM-FLMM-full}
\begin{figure}[H]
		\centering
		\includegraphics[width=0.55\linewidth]{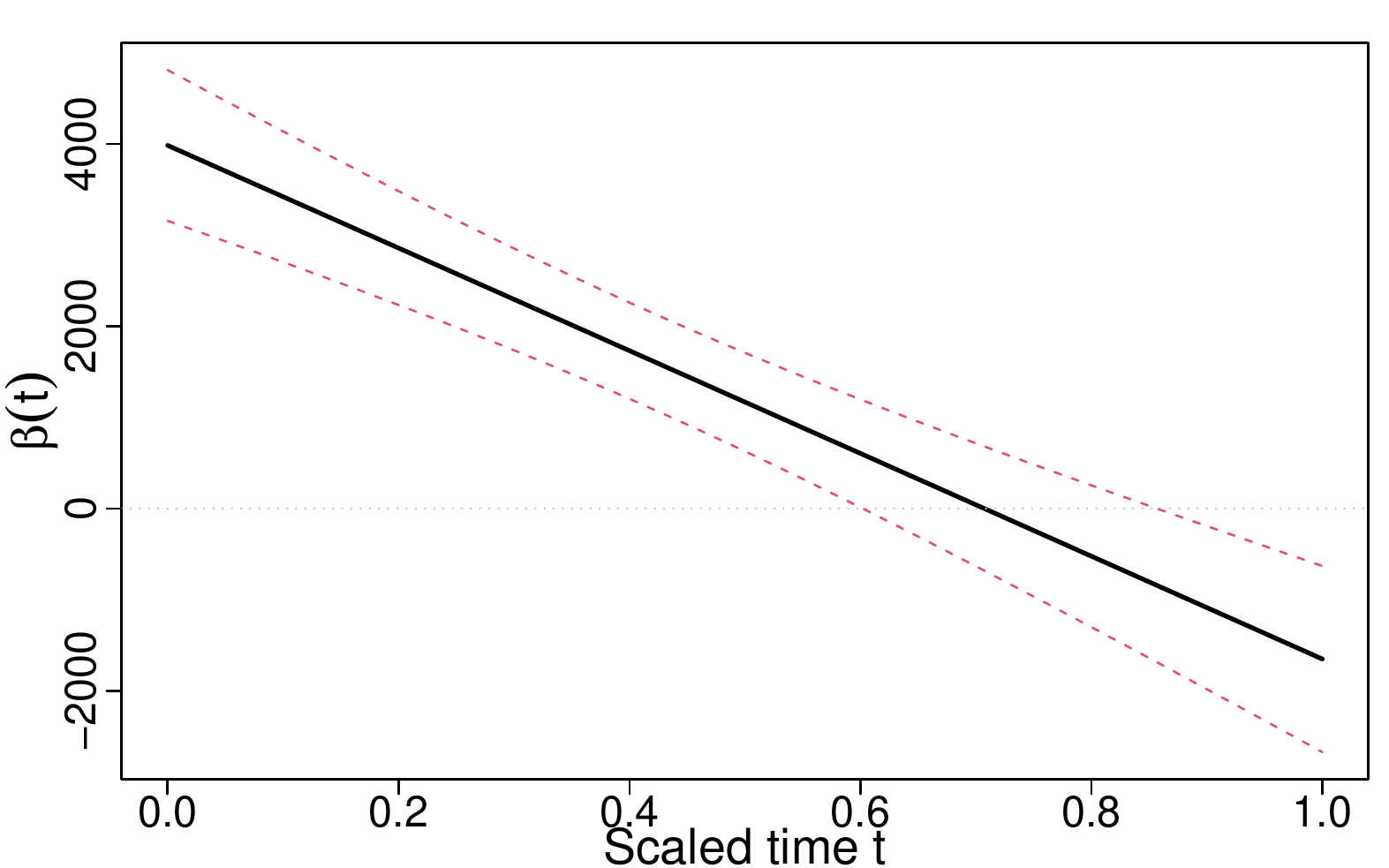}
		\caption{Estimated coefficient function of the effect from scaled VDC on EOD time, $\beta(t)$. Mean estimate is in the solid curve and dashed curves indicated the $95\%$ pointwise confidence intervals.}
		\label{fig:betat}
	\end{figure}

\begin{figure}[H]
		\centering
		\includegraphics[width=1\linewidth]{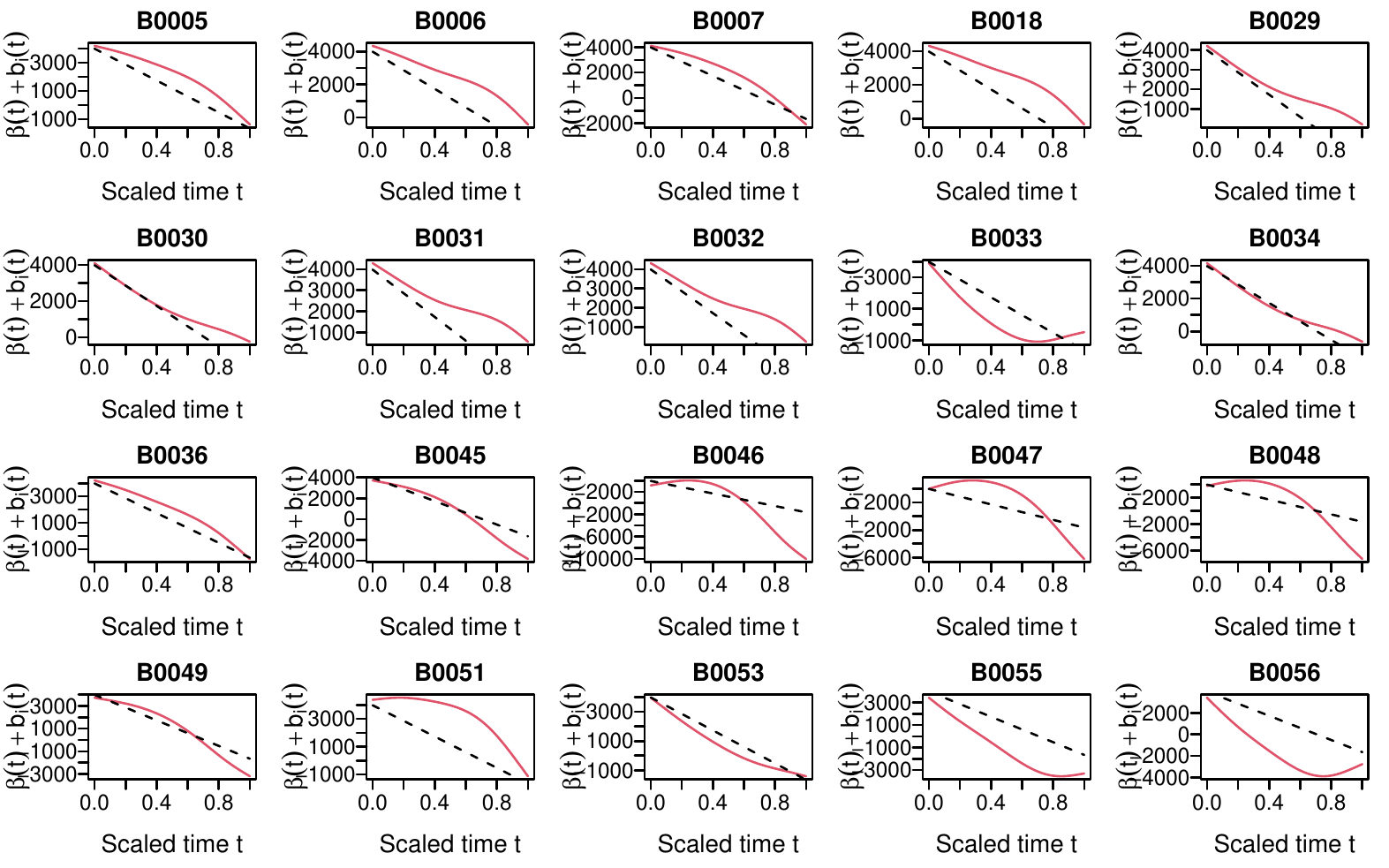}
		\caption{Estimated functional coefficients of scaled VDC by battery from FLMM model predicting EOD time. Dashed curve represents overall coefficient function $\hat{\beta}(t)$. Battery-specific coefficient functions, $\hat{\beta}(t) + \hat{b}_i(t)$, are depicted by solid curves.}
		\label{fig:bt}
	\end{figure}

	The FDM-FLMM for EOD $b_{ic}$ considered in Section \ref{ss:da_fulldata} is
	\begin{multline}\label{eq:all_flmm}
		b_{ic} = \alpha_0+w_{0i} + (\alpha_1+w_{1i})c + \alpha_2b_{i,c-1} + \alpha_3\exp(-1/\Delta_{ic}) \\
  + \int_0^1 \left(\beta(t) + b_i(t)\right)x_{ic}(t)dt+\epsilon_{ic}.
	\end{multline}
	The estimations for fixed coefficients and their se are $\hat{\alpha}_0 = -3845.733$ with se 931.087, $\hat{\alpha}_1 = -1.687$ with se 0.383, $\hat{\alpha}_2 = 0.590$ with se 0.014, and $\hat{\alpha}_3 = 546.725$ with se 25.225. One can see that the coefficients estimated in \eqref{eq:all_lme} and \eqref{eq:all_flmm} for identical covariates are similar. Figure \ref{fig:betat} illustrates the estimated coefficient function $\beta(t)$ of the effect of scaled VDCs on EOD plotted point-wise and Figure \ref{fig:bt} shows the 20 battery-specific effect from scaled VDCs on EOD, $\beta(t) + b_i(t)$.
 
	From the form of \eqref{eq:all_flmm}, the effect from the scaled VDCs is a weighted average of the scaled VDC with the weight set by the coefficient function $\beta(t)$ adjusted with $b_i(t)$ for each battery. The estimated $\beta(t)$ has the shape close to a straight line with negative slope and sign of values switch at around $0.7$. Recall the shape of a typical scaled VDCs (Figure \ref{fig:fpcs}), we can see that this term $\int_0^1 \left(\beta(t) + b_i(t)\right)x_{ic}(t)dt$ acts as an adjustment to EOD based on a weighted average of scaled VDCs. Without battery specific effect $b_i(t)$, this adjustment is similar to an area under the curve. Depending on the battery, the variations in the shapes of scaled VDCs help set the weight to be $\beta(t) + b_i(t)$.
	
\end{appendix}

\end{document}